\documentclass[12pt]{article}

\usepackage{amsmath,amsthm,amsfonts,amssymb}
\usepackage{graphicx}
\usepackage{a4wide}

\usepackage{authblk}

\title{A Markov chain model to investigate the spread of antibiotic-resistant bacteria in hospitals}

\date{\today}

\author[1]{Fabio A.C.C. Chalub}

\author[2]{Antonio G\'{o}mez-Corral}

\author[3]{Mart\'{\i}n L\'{o}pez-Garc\'{\i}a}

\author[4]{F\'{a}tima Palacios-Rodr\'{\i}guez}

\affil[1]{Departamento de Matem\'{a}tica \& Centro de Matem\'{a}tica e Aplica\c{c}\"{o}es, Faculdade de Ci\^{e}ncias e Tecnologia, Universidade Nova de Lisboa, Quinta da Torre, 2829-516 Caparica, Portugal}

\affil[2]{Department of Statistics and Operations Research, Complutense University of Madrid, 28040-Madrid, Spain}

\affil[3]{Department of Applied Mathematics, School of Mathematics, University of Leeds, LS2 9JT Leeds, United Kingdom}

\affil[4]{Department of Statistics and Operations Research, Faculty of Mathematics, University of Seville, Calle Tarfia s/n, 41012-Seville, Spain}

\begin{document}

\maketitle

\begin{abstract}
Ordinary differential equation (ODE) models used in mathematical epidemiology assume explicitly or implicitly large populations. For the study of infections in a hospital this is an extremely restrictive assumption as typically a hospital ward has a few dozen, or even fewer, patients. This work reframes a well-known model used in the study of the spread of antibiotic-resistant bacteria in hospitals, to consider the pathogen transmission dynamics in small populations. In this vein, this paper proposes a Markov chain model to describe the spread of a single bacterial species in a hospital ward where patients may be free of bacteria or may carry bacterial strains that are either sensitive or resistant to antimicrobial agents. We determine the probability law of the \emph{exact} reproduction number ${\cal R}_{exact,0}$, which is here defined as the random number of secondary infections generated by those patients who are accommodated in a predetermined bed before a patient who is free of bacteria is accommodated in this bed for the first time. Specifically, we decompose the exact reproduction number ${\cal R}_{exact,0}$ into two contributions allowing us to distinguish between infections due to the sensitive and the resistant bacterial strains. Our methodology is mainly based on structured Markov chains and the use of related matrix-analytic methods. This guarantees the compatibility of the new, finite-population model, with large population models present in the literature and takes full advantage, in its mathematical analysis, of the intrinsic stochasticity.
 \end{abstract}

\textbf{keywords:}
Epidemic model, Markov chain, quasi-birth-death process, reproduction number
%

\maketitle

\textbf{Abbreviations:} LD-QBD, level-dependent quasi-birth-death; SI, susceptible-infective; SIS, susceptible-infective-susceptible; SIR, susceptible-infective-removed.

\renewcommand\thefootnote{\fnsymbol{footnote}}
\setcounter{footnote}{1}

\section{Introduction}
\label{sec:1}
\vspace*{12pt}
Nosocomial infections caused by antibiotic (or antimicrobial) resistant bacteria ---such as methicillin-resistant \emph{Staphylococcus aureus} (Haaber et al. \cite{Haaber17}), multidrug-resistant \emph{Mycobacterium tuberculosis} (Gygli et al. \cite{Gygli17}), vancomysin-resistant \emph{Enterococci} (Miller et al. \cite{Miller14}), and multidrug-resistant \emph{Gram-negative bacilli} (Breijyeh et al. \cite{Breijyeh20}), among others--- are usually most prevalent in intensive care units and hospital settings where patients are susceptible to the acquisition of carriage, mainly due to high selective antibiotic pressure or frequent opportunities for cross-transmission. Compared to infections caused by antibiotic sensitive bacteria, infections caused by resistant bacteria drastically reduce the probability of successfully treating bacterial infections, prolong hospitalizations, and increase health-care costs, morbidity and mortality, among other implications; see e.g. D'Agata et al. \cite{D'Agata07} and references therein. The collaborative paper \cite{Murray22} is a first comprehensive assessment of the global burden of antimicrobial resistance and an evaluation of the availability of data in 2019.
\par To examine the implications of the emergence and spread of antibiotic resistance, mathematical modelling (Niewiadomska et al. \cite{Niewiadomska19}) provides a platform for {\it in silico} experiments that improve our ability to determine the quantitative effects of the transmission process and potential control measures. Most of the existing models follow a deterministic approach, mostly based on the use of ordinary differential equations, on either within-host (Techitnutsarut and Chamchod \cite{Techitnutsarut21}) or between-host (Bagkur et al. \cite{Bagkur22}; D'Agata et al. \cite{D'Agata02}; Lipsitch et al. \cite{Lipsitch2000}) frameworks; for a novel work formulating a two-level population model, we refer the reader to the paper by Webb et al. \cite{Webb05}. An excellent summary on antibiotic-resistance modelling is the review of Spicknall et al. \cite{Spicknall13}, where the peer-reviewed literature on between-host resistance modelling ---in particular, papers published from 1993 to 2011--- is categorized by classifying each paper's model structure into up to six categories based on the underlying inherent assumptions. In the probabilistic setting, Seigal et al. \cite{Seigal2017} introduce a transmission model ---which uses the negative binomial distribution---, present a statistical hypothesis test that calculates the significance of resistance trends occurring in a hospital, and apply the method to each of sixteen antibiotics in a case study of spectrum $\beta$-lactamases samples collected from patients at a community hospital over a 2.5-year period.
\par In this paper, we complement the work of G\'{o}mez-Corral and L\'{o}pez-Garc\'{\i}a \cite[Section 3.3]{GC18}, which is related to a stochastic version of the deterministic between-host model of Lipsitch et al. \cite{Lipsitch2000} for antimicrobial resistance in nosocomial pathogens; in a more general context, see the book of Allen \cite{Allen03} for a comprehensive discussion of results on deterministic epidemic models and their stochastic counterparts. In order to further clarify differences between the sensitive and the resistant bacterial strains, our objective here is to characterize the probability law of the \emph{exact} reproduction number ${\cal R}_{exact,0}$ (Artalejo and L\'{o}pez-Herrero \cite{Artalejo13}; G\'{o}mez-Corral et al. \cite{GC23}) by decomposing this number into two random contributions ${\cal R}_{exact,0}^S$ and ${\cal R}_{exact,0}^R$ that count secondary infections due to the sensitive and the resistant bacteria, respectively. The main features of the exact reproduction number ${\cal R}_{exact,0}$, compared to the basic reproduction number ${\cal R}_0$ in \cite{Lipsitch2000} (i.e., its deterministic counterpart), are inherently linked to the fact that it relates here to a \emph{marked} bed ---initially accommodating an inpatient colonized with bacteria--- instead of a single infective inpatient, it eliminates the effect of repeated infectious contacts, and it is not necessarily defined at the time of invasion, but at any later time. Therefore, the expected value of ${\cal R}_{exact,0}$ can be thought of as a more accurate index than the value ${\cal R}_0$, especially in the setting of a hospital ward \cite{Lipsitch2000} and an early stage of the epidemic, and the marginal mass functions of ${\cal R}_{exact,0}^S$ and of ${\cal R}_{exact,0}^R$ can be used to predict the prevalence of nosocomial infections in terms of the fitness cost of resistance to antimicrobial agents, and explore how specific interventions based on admitting more patients already colonized with sensitive bacteria will prevent the transmission of the resistant strain within the ward, but increase the transmission of the sensitive one.
\par The analysis in \cite[Section 3.3]{GC18} illustrates how to apply a perturbation approach of finite level-dependent quasi-birth-death (LD-QBD) processes to two-strain susceptible-infective (SI) and susceptible-infective-susceptible (SIS) epidemic models. Specifically, the random length of an outbreak, the final size of the epidemic, the peak of infection and the state of the population at an arbitrary time in these epidemic models are analyzed in \cite{GC18} as first-passage times, hitting probabilities, extreme values and stationary regime, respectively, in the underlying LD-QBD process. See, e.g., the papers by De Nitto Person\`{e} and Grassi \cite{DeNitto96}, Gaver et al. \cite{Gaver84}, and G\'{o}mez-Corral et al. \cite{GC20} for a detailed discussion on LD-QBD processes and their applications in the context of varicella-zoster virus infections. The work to be presented here is part of our ongoing study on the use of Markov chains, including LD-QBD processes, and related matrix-analytic methods in a variety of stochastic epidemic models, such as SIS and SIR models with two strains and cross-immunity (Almaraz and G\'{o}mez-Corral \cite{Almaraz19}; Amador et al. \cite{Amador19}), discrete and continuous versions of SIS models (Chalub and Sousa \cite{Chalub14}; G\'{o}mez-Corral et al. \cite{GC21}), quarantine of hosts (Amador and G\'{o}mez-Corral \cite{Amador20}), limited resources in epidemics (Amador and L\'{o}pez-Herrero \cite{Amador18}) and  vaccination strategies (Fern\'{a}ndez and G\'{o}mez-Corral \cite{Fernandez21}; Gamboa and L\'{o}pez-Herrero \cite{Gamboa22}).
\par This article is organized as follows. Section 2 provides the mathematical description of a Markov chain model for the potential spread of a single bacterial species in a hospital ward where patients are accommodated in beds and may either be free of bacteria or carry antibiotic-sensitive or resistant bacteria. In particular, the model is first formulated as a LD-QBD process and then related to the deterministic model of Lipsitch et al. \cite{Lipsitch2000}. In Section 3 the propagation potential of the bacterial strains in early stages of the outbreak is studied in terms of suitably defined versions ${\cal R}_{exact,0}^S$ and ${\cal R}_{exact,0}^R$ of the \emph{exact} reproductive number for the sensitive and the antibiotic-resistant bacterial strains, when the focus is on infections generated from a predetermined bed and an invasion time. In Sections 4 and 5, a discussion of numerical experiments and concluding remarks are presented.
\section{Mathematical model description: a Markov chain model}
\label{sec:2}
\vspace*{12pt}
We consider a stochastic version of the SIS model in \cite{Lipsitch2000} for the transmission dynamics of two bacterial strains with partial cross-immunity in a hospital ward. Patients in this ward are accommodated in $N$ beds, and may carry bacterial strains that are antibiotic-resistant (and possibly also antibiotic-sensitive bacteria) or only antibiotic-sensitive to a first antimicrobial agent, referred to as drug 1, or they may be free of bacteria. After a certain time, within-host dynamics will lead to the elimination of one strain. The result will depend on the selection pressures in the host, whence we assume that there is no coinfection, cf. Lipsitch et al. \cite{Lipsitch2000,Lipsitch09}. Resistance to a second antimicrobial agent, referred as drug 2, is not present in the bacteria. Despite the tendency to reduce the use of antibiotics, there are medically relevant situations, in particular in prophylaxis against surgical site infections, after a bite or wound that could get infected, or if the patient has a higher risk of infection, in which antibiotics are administrated irrespectively of external signs of infection, cf. Calderwood et al.  \cite{Calderwood23}, Pak et al. \cite{Pak21} and the NHS-UK
standards\footnote{https://www.nhs.uk/conditions/antibiotics/uses/, consulted at 07/12/2023.}. The model in \cite{Lipsitch2000} assumes that both antimicrobial agents are administrated in a prophylactic manner, which means that patients routinely receive drugs 1 and 2 at a rate which does not depend on whether or not they are colonized with bacteria; more concretely, treatment with drug 1 clears carriage of sensitive bacteria at rate $\tau_1$ per day, and treatment with drug 2 clears carriage of both bacterial strains at rate $\tau_2$ per day. Mutation from the sensitive to the resistant bacterial strain is not possible during the timescales of the outbreak, nor vice versa.
\begin{figure}
	\centering
	\begin{center}
        \includegraphics[width=0.9\textwidth]{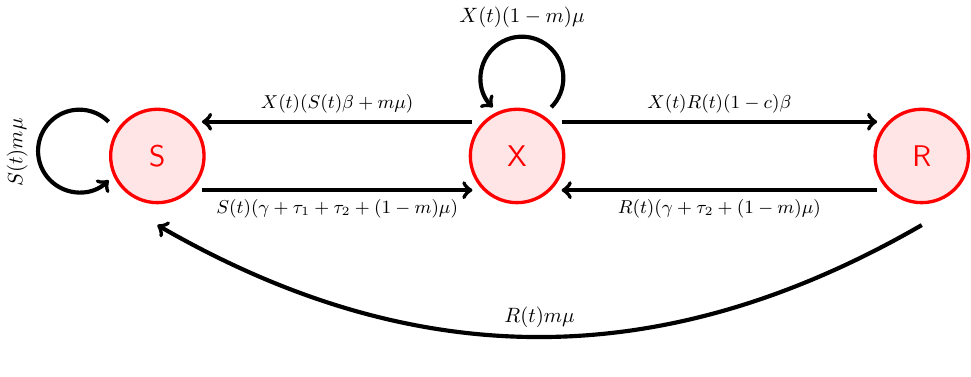}
	\end{center}
	\caption{Diagram of transitions among compartments. The factors in each arrow indicate the transition rate per unit of time from one compartment to the next one. $S(t)$, $R(t)$ and $X(t)$ are the number of individuals in classes S, R, and X at time $t$, i.e., the number of individuals infected only with the antibiotic-sensitive strain, the antibiotic-resistant strain, and non-infected individuals, respectively. Model parameters are described in Table~\ref{table:parameters}.}
	\label{fig:diagram}
\end{figure}
\begin{table}
    \centering
    \begin{tabular}{l|l}
        \hline\noalign{\smallskip}
        $\beta$         & per capita transmission rate \\
        $m$             & fraction of new admissions carrying the sensitive bacterial strain \\
        $\mu$           & discharge rate, equal to admission rate \\
        $c$             & fitness cost of resistance to drug 1 \\
        $\gamma$        & rate of spontaneous clearance \\
        $\tau_1$   & rate of clearance due to drug 1 \\
        $\tau_2$   & rate of clearance due to drug 2 \\
        \noalign{\smallskip}\hline
    \end{tabular}
    \caption{Parameters of the model.}
    \label{table:parameters}
\end{table}
\par Bacteria may be transmitted between patients via direct contacts, which turns a patient who is free of bacteria into colonized with either sensitive bacteria at rate $\beta$ per day, or resistant bacteria at rate $(1-c)\beta$ per day, where $\beta$ is the per capita infection rate and $c\in (0,1)$ is the fitness cost of resistance to drug 1. Patients are assumed to be admitted by and discharged from the hospital ward\footnote{Hereinafter we will use the term \emph{discharge from the hospital unit} to refer to the abandonment of the hospital ward due to any cause, such as the transfer of the patient to his home or another hospital ward, and his possible death.} at rate $\mu$ per day, in such a way that they are replaced instantly by new patients who either are colonized with sensitive bacteria or are free of bacteria with proportion $m$ and $1-m$, respectively, where $m$ amounts to the proportion of people colonized with sensitive bacteria in the population at large. Spontaneous clearance of bacteria is seen to occur at rate $\gamma$ per day.
\par Under the assumption of exponentially distributed sojourn times, the state of the hospital ward may be captured by means of a time-homogeneous continuous-time Markov chain ${\cal X}=\{(S(t)+R(t),R(t)): t\geq 0\}$, where $S(t)$ and $R(t)$ record the number of patients colonized with sensitive and resistant bacteria, respectively, at time $t$. This results in the number $X(t)=N-S(t)-R(t)$ of patients who are free of bacteria. The bivariate process ${\cal X}$ can be seen as a LD-QBD process taking values in the finite set ${\cal S}=\cup_{i=0}^N l(i)$ with \emph{levels} $l(i)=\{(i,j): j\in\{0,...,i\}\}$, for integers $i\in\{0,...,N\}$. To be concrete, the infinitesimal dynamics of ${\cal X}$ are governed by the following non-vanishing transition rates from state $(i,j)$ to state $(i',j')$:
\begin{eqnarray*}
    q_{(i,j),(i',j')} &=& \left\{\begin{array}{ll}
        (N-i)((i-j)\beta+m\mu), & \hbox{if $(i',j')=(i+1,j)$,}
        \\
        (N-i)j(1-c)\beta, & \hbox{if $(i',j')=(i+1,j+1)$,}
        \\
        jm\mu, & \hbox{if $(i',j')=(i,j-1)$,}
        \\
        j(\gamma+\tau_2+(1-m)\mu), & \hbox{if $(i',j')=(i-1,j-1)$,}
        \\
        (i-j)(\gamma+\tau_1+\tau_2+(1-m)\mu), & \hbox{if $(i',j')=(i-1,j)$,}
        \end{array}\right.
\end{eqnarray*}
for states $(i,j), (i',j')\in {\cal S}$ with $(i',j')\neq (i,j)$. Clearly, $q_{(i,j)(i,j+1)}=0$. The transitions above represent, respectively, $X\to S$, $X\to R$, $R\to S$, $R\to X$, $S\to X$, and $S\to R$; see Figure~\ref{fig:diagram} for further details. Furthermore $q_{(i,j),(i,j)}=-q_{(i,j)}$, where
\begin{eqnarray*}
    q_{(i,j)} &=& (N-i)((i-jc)\beta+m\mu)+(i-j)(\gamma+\tau_1+\tau_2+(1-m)\mu)+j(\gamma+\tau_2+\mu).
\end{eqnarray*}
\par In the above description, two \emph{hidden} events for process ${\cal X}$ can occur due to the replacement of a patient who is discharged from the hospital ward by a newly admitted patient who belongs to the same compartment; i.e., the process ${\cal X}$ does not allow us to record a transition from state $(i,j)$ to $(i,j)$, which occurs with rate $(N-i)(1-m)\mu+(i-j)m\mu$. This corresponds to a change in the time scale, without further effects in the final states and associated probability.
\subsection{The Markov chain model \emph{versus} its deterministic counterpart}
\label{subsec:21}
\vspace*{12pt}
The proposed Markov chain model is a stochastic realization of the deterministic model introduced in~\cite{Lipsitch2000}. In that case, it is considered the set of ordinary differential equations
\begin{eqnarray*}
    s' &=& m\mu+\frac{\beta sx}{s+r+x}-(\gamma+\tau_1+\tau_2+\mu)s,
    \\
    r' &=& \frac{(1-c)\beta rx}{s+r+x}-(\gamma+\tau_2+\mu)r,
    \\
    x' &=& (1-m)\mu+(\gamma+\tau_1+\tau_2)s+(\gamma+\tau_2)r - \frac{\beta sx}{s+r+x}-\frac{(1-c)\beta rx}{s+r+x}-\mu x,
\end{eqnarray*}
where $s(t)$, $r(t)$, and $x(t)$ represent the proportion of patients colonized with sensitive and resistant bacteria, and the fraction of patients free of bacteria, respectively, at time $t$, and $\beta$ is the transmission rate. We did not assume \emph{a priori} the normalization, for reasons that will be clarified in the sequel. The normalization, however, follows from the simple fact that $(s+r+x)'=\mu(1-s-r-x)$, and, therefore, we may assume $s(t)+r(t)+x(t)=1$, for all $t\ge 0$, and omit one of the equations.
\par Defining $z=s+r$ the above system is equivalent to
\begin{eqnarray}
    \label{eq:Lipsith1}
    z' &=& m\mu+\beta (z-cr)(1-z)-(\gamma+\tau_1+\tau_2+\mu)z+\tau_1r ,\\
    \label{eq:Lipsith2}
    r' &=& r\Big((1-c)\beta (1-z)-(\gamma+\tau_2+\mu)\Big).
\end{eqnarray}
\par It is possible to prove, in a very precise way, that the Markov chain introduced in the present manuscript converges in the large population limit to the model~(\ref{eq:Lipsith1})--(\ref{eq:Lipsith2}). More precisely, we show that the large population limit of the Markov chain is a first-order partial differential equation, see equation below, such that its characteristics are the trajectories of the solutions of the system (\ref{eq:Lipsith1})--(\ref{eq:Lipsith2}). The proof follows ideas from~\cite{Chalub09}; see also~\cite{Chalub11,Chalub14}.
\par Namely, for the Markov chain, we consider the \emph{master equation}
\begin{eqnarray*}
    p\left(\frac{i'}{N},\frac{j'}{N},t+\Delta t\right) &=& p\left(\frac{i'}{N},\frac{j'}{N},t\right)+\sum_{(i,j)\ne(i',j')}q_{(i,j)(i',j')}p\left(\frac{i}{N},\frac{j}{N},t\right)\Delta t,
\end{eqnarray*}
where $p(z,r,t)$ is the probability to find the Markov chain at state $(\lfloor Nz\rfloor,\lfloor Nr\rfloor)$ at time $t$. In the weak formulation, the last equation reads
\begin{eqnarray*}
    \sum_{(i',j')}p\left(\frac{i'}{N},\frac{j'}{N},t+\Delta t\right)\varphi\left(\frac{i'}{N},\frac{j'}{N}\right) &=&  \sum_{(i,j)}p\left(\frac{i}{N},\frac{j}{N},t\right)\sum_{(i',j')\ne(i,j)}q_{(i,j),(i'j')}\varphi\left(\frac{i'}{N},\frac{j'}{N}\right),
\end{eqnarray*}
where $\varphi$ is an adequate test function, i.e., $\varphi(z+\Delta z,r\pm \Delta r) = \varphi(z,r)+\Delta z\partial_z\varphi(z,r)+\Delta r\partial_r\varphi(z,r)+ o(\Delta z,\Delta r)$. Therefore, it is seen that
{\footnotesize
\begin{eqnarray*}
    \sum_{(i',j')}q_{(i,j)(i',j')}\varphi\left(\frac{i'}{N},\frac{j'}{N}\right) &=& \varphi\left(\frac{i}{N},\frac{j}{N}\right)+\Big((N-i)\Big((i-j)\frac{\beta}{N}+m\mu+j(1-c)\beta\Big)-j(\gamma+\tau_2+\gamma+(1-m)\mu)\\
    & & -(i-j)(\gamma+\tau_1+\tau_2+(1-m)\mu)\Big)\frac{\Delta t}{N}\partial_z\varphi\left(\frac{i}{N},\frac{j}{N}\right)\\
    & & +\Big((N-i)\frac{\beta}{N}(1-c)-\mu m-(\tau_2+\gamma+\mu(1-m))\Big)\frac{j\Delta t}{N}\partial_r\varphi\left(\frac{i}{N},\frac{j}{N}\right)+o\left(\frac{\Delta t}{N}\right)\\
    &=& \varphi\left(z,r\right)+\Delta t\Big(\Big((1-z)\beta(z-cr)+\mu m-z(\gamma+\tau_1+\tau_2+\mu)+r\tau_1\Big)\partial_z\varphi(z,r)\\
    & & +\Big((1-z)\beta(1-c)-\tau_2+\gamma+\mu)\Big)r\partial_r\varphi(z,r) + o\left(\frac{1}{N}\right)\Big).
\end{eqnarray*}
}
\par After taking the limit $\Delta t\to0$ and $N\to\infty$, we conclude that probabilities $p(z,r,t)$ satisfy the equation
\begin{eqnarray*}
    \partial_t p &=& -\partial_z\Big((1-z)\beta(z-cr)+\mu m-z(\gamma+\tau_1+\tau_2+\mu)+r\tau_1\Big)-\partial_r \Big((1-z)\beta(1-c)-(\tau_2+\gamma+\mu)\Big) .
\end{eqnarray*}
The characteristics of this equation are the solutions of the system $(z,r)$ defined above. Note the relevance of not assuming \emph{a priori} the normalization $s+r+x=1$, as the $1/N$ normalization in the transmission rate is explicitly used in order to have a well-defined limit.
\\ \\
{\bf Remark 1:} {\it There is no specific reason to stop the Taylor expansion of the Markov chain model at the first order in $1/N$; in fact, the second order term will model random effects. The idea is clear in population genetic models (cf.~\cite{Chalub09,Chalub14b}); in epidemiological models, the interpretation of the diffusion coefficient is not clear~\cite{Chalub14}. The resulting equation is a partial differential equation with degenerated coefficient and its rigorous mathematical analysis presents serious challenges~\cite{Chugonova23,Danilkina18}.}
\section{The exact reproduction number ${\cal R}_{exact,0}$}
\label{sec:3}
Let us consider an invasion time, i.e., $S(0)+R(0)=1$, and assume that the initially infected patient is accommodated in a \emph{marked} bed. It is likely that the initially infected patient may be discharged from the hospital ward before recovering, in which case the newly admitted patient who settles into the marked bed instead may either be free of bacteria or colonized with sensitive bacteria. In the latter case, this means that the bacteria remains present in the hospital ward, with the marked bed being a source of bacterial transmission. Thus, the dynamics of bacterial dissemination during an early stage of the epidemic are not linked only to the initially infected individual, but to those patients who settle one after another in the marked bed.
\par We define here the exact reproduction number ${\cal R}_{exact,0}$ as the number of infections generated by the patients who are accommodated ---one after another--- in the marked bed before one of them is free of bacteria. The most remarkable feature of process ${\cal X}$ is the transmission of sensitive and resistant bacteria, whence we express the exact reproduction number ${\cal R}_{exact,0}$ in terms of two random contributions ${\cal R}_{exact,0}^S$ and ${\cal R}_{exact,0}^R$ according to the fact that infections are due to the sensitive bacterial strain and the resistant bacterial strain, respectively.
\\ \\
{\bf Remark 2:} {\it In Ref. \cite{Lipsitch2000}, the basic reproduction number\footnote{In the terminology of \cite{Lipsitch2000}, the parameter $\beta$ is the transmission rate, which is equivalent to $N\beta$ if $\beta$ denotes the per capita infection rate.}
\begin{eqnarray*}
    R_0 &=& \frac{\beta}{\gamma+\tau_2+\mu}
\end{eqnarray*}
is linked to a single patient colonized with resistant bacteria, who is hospitalized in a hypothetical hospital ward where all other inpatients entered uncolonized; i.e., secondary cases contributing to $R_0$ are resistant infections generated by this initially colonized patient before either becoming free of bacteria or leaving the hospital ward. An interesting question, which will be addressed in Section \ref{sec:4}, concerns the relationship between the expected number $E[{\cal R}_{exact,0}^R | (S(0)+R(0),R(0))=(1,1)]$ and its deterministic counterpart $R_0$.}
\\
\par In this section, our objective is to determine the joint probability law of $({\cal R}_{exact,0}^S,{\cal R}_{exact,0}^R)$ by evaluating the conditional probabilities
\begin{eqnarray*}
    & P\left( \left.({\cal R}_{exact,0}^S,{\cal R}_{exact,0}^R)=(s,0) \right| (S(0)+R(0),R(0))=(1,0)\right), &
    \\
    & P\left( \left.({\cal R}_{exact,0}^S,{\cal R}_{exact,0}^R)=(s,r) \right| (S(0)+R(0),R(0))=(1,1)\right), &
\end{eqnarray*}
for integers $s,r\in\mathbb{N}_0$. In evaluating these probabilities, we first denote the \emph{status} of the patient who is accommodated in the marked bed at time $t$ by  $B(t)$ ---in such a way that $B(t)=0$ if the patient is free of bacteria, and $B(t)=1_S$ and $1_R$ if the patient is colonized with sensitive and resistant bacteria, respectively---, and we define the following more general conditional probabilities:
\begin{itemize}
    \item[{(i)}] For states $(i,j)\in {\cal S}$ with $i\in\{1,...,N\}$ and $j\in\{0,...,i-1\}$, we consider
        \begin{eqnarray*}
            P_{(i,j),1_S}(s,r) &=& P\left(\left.({\cal R}_{exact,0}^S,{\cal R}_{exact,0}^R)=(s,r) \right| (S(0)+R(0),R(0))=(i,j), B(0)=1_S\right),
        \end{eqnarray*}
        for $s,r\in\mathbb{N}_0$. Clearly, $P_{(i,j),1_S}(s,r)=0$ if $r\in\mathbb{N}$, for $i\in\{1,...,N\}$ and $j\in\{0,...,i-1\}$.
    \item[{(ii)}] For states $(i,j)\in {\cal S}$ with $i\in\{1,...,N\}$ and $j\in\{1,...,i\}$, we consider
        \begin{eqnarray*}
            P_{(i,j),1_R}(s,r) &=& P\left(\left.({\cal R}_{exact,0}^S,{\cal R}_{exact,0}^R)=(s,r) \right| (S(0)+R(0),R(0))=(i,j), B(0)=1_R\right),
        \end{eqnarray*}
        for $s,r\in\mathbb{N}_0$.
\end{itemize}
\subsection{The special case $B(0)=1_S$}
\label{subsec:1}
In this subsection, we derive an iterative procedure that, starting from the family of conditional probabilities $\{P_{(i,j),1_S}(0,0): i\in\{1,...,N\}, j\in\{0,...,i-1\}\}$, evaluates the family of probabilities $\{P_{(i,j),1_S}(s,0): i\in\{1,...,N\}, j\in\{0,...,i-1\}\}$ in terms of the probabilities in $\{P_{(i,j),1_S}(s-1,0): i\in\{1,...,N\}, j\in\{0,...,i-1\}\}$, for $s\in\mathbb{N}$.
\par We use first-step analysis to obtain the system of linear equations
{\footnotesize
\begin{eqnarray}
    P_{(i,0),1_S}(0,0) &=& \frac{\gamma+\tau_1+\tau_2+(1-m)\mu}{q_{(i,0)}}+\frac{(i-1)(\gamma+\tau_1+\tau_2+(1-m)\mu)}{q_{(i,0)}}P_{(i-1,0),1_S}(0,0) \nonumber
    \\
    & & +\left(\frac{(i-1)m\mu}{q_{(i,0)}}+\frac{(N-i)(1-m)\mu}{q_{(i,0)}}\right) P_{(i,0),1_S}(0,0)+\frac{(N-i)((i-1)\beta+m\mu)}{q_{(i,0)}} P_{(i+1,0),1_S}(0,0),
    \label{eq:1}
    \\
    P_{(i,0),1_S}(s,0) &=& \frac{m\mu}{q_{(i,0)}} P_{(i,0),1_S}(s-1,0)+\frac{(N-i)\beta}{q_{(i,0)}} P_{(i+1,0),1_S}(s-1,0)+\frac{(i-1)(\gamma+\tau_1+\tau_2+(1-m)\mu)}{q_{(i,0)}} P_{(i-1,0),1_S}(s,0) \nonumber
    \\
    & & +\left(\frac{(i-1)m\mu}{q_{(i,0)}}+\frac{(N-i)(1-m)\mu}{q_{(i,0)}}\right)P_{(i,0),1_S}(s,0)+\frac{(N-i)((i-1)\beta+m\mu)}{q_{(i,0)}} P_{(i+1,0),1_S}(s,0), \label{eq:2}
\end{eqnarray}
}
for integers $i\in\{1,...,N\}$ and $s\in\mathbb{N}$. Equations (\ref{eq:1}) and (\ref{eq:2}) can be readily written as a single equation in matrix form by using column vectors ${\bf P}_{0,1_S}(s)$, for $s\in\mathbb{N}_0$, and ${\bf p}_{0,1_S}$ with $i$-th entries $P_{(i,0),1_S}(s,0)$ and $q^{-1}_{(i,0)}(\gamma+\tau_1+\tau_2+(1-m)\mu)$, respectively, for $i\in\{1,...,N\}$. Specifically, it is seen that
\begin{eqnarray}
    {\bf P}_{0,1_S}(s) &=& \left({\bf I}_N-{\bf C}_{0}(0)\right)^{-1}\Big(\delta_{s,0}{\bf p}_{0,1_S}+(1-\delta_{s,0}){\bf B}_{0}{\bf P}_{0,1_S}(s-1)\Big),
    \label{eq:3}
\end{eqnarray}
where ${\bf I}_a$ denotes the identity matrix of order $a$, $\delta_{a,b}$ represents the Kronecker delta, and ${\bf B}_{0}$ and ${\bf C}_{0}(0)$ are suitably defined matrices of coefficients; see Appendix \ref{app:A}.
\par For initial states $(i,j)\in {\cal S}$ with $i\in\{j+1,...,N\}$ and $j\in\{1,...,N-1\}$, a similar approach leads us to the following equalities for the column vectors ${\bf P}_{j,1_S}(s)$, for $j\in\{1,...,N-1\}$ and $s\in\mathbb{N}_0$, with $i$-th entry $P_{(j+i,j),1_S}(s,0)$, for $i\in\{1,...,N-j\}$:
\begin{eqnarray}
\nonumber
    {\bf P}_{j,1_S}(s) &=& \delta_{s,0}{\bf p}_{j,1_S}+(1-\delta_{s,0}){\bf B}_{j}{\bf P}_{j,1_S}(s-1)+{\bf C}_{j-1}(1){\bf P}_{j-1,1_S}(s)\\
    &&\quad+(1-\delta_{j,N-1})\left( {\bf C}_{j}(0){\bf P}_{j,1_S}(s)+{\bf C}_{j+1}(2){\bf P}_{j+1,1_S}(s)\right),
    \label{eq:4}
\end{eqnarray}
where the column vector ${\bf p}_{j,1_S}$ has $i$-entry $q^{-1}_{(j+i,j)}(\gamma+\tau_1+\tau_2+(1-m)\mu)$, for $i\in\{1,...,N-j\}$, and ${\bf B}_{j}$, ${\bf C}_{j}(0)$, ${\bf C}_{j-1}(1)$ and ${\bf C}_{j+1}(2)$ are matrices of coefficients; see Appendix \ref{app:B}.
\par For a fixed value $s\in\mathbb{N}_0$, Eq. (\ref{eq:4}) can be seen as a tridiagonal-by-blocks system of linear equations for the unknown vectors ${\bf P}_{j,1_S}(s)$, for $j\in\{1,...,N-1\}$, which can be solved using block-Gaussian elimination in terms of previously evaluated vectors ${\bf P}_{j,1_S}(s-1)$.
\\ \\
{\bf Theorem 1} {\it For $s\in\mathbb{N}_0$, the column vectors in $\{{\bf P}_{j,1_S}(s): j\in\{1,...,N-1\}\}$ satisfy the recurrence equations
\begin{eqnarray}
    {\bf P}_{j,1_S}(s) &=& {\bf h}_{j,1_S}(s)+(1-\delta_{j,N-1}){\bf H}^{-1}_{j,1_S}{\bf C}_{j+1}(2){\bf P}_{j+1,1_S}(s),
    \label{eq:5}
\end{eqnarray}
with
\begin{eqnarray*}
    {\bf h}_{j,1_S}(s) &=& {\bf H}^{-1}_{j,1_S}\Big(\delta_{s,0}{\bf p}_{j,1_S}+(1-\delta_{s,0}){\bf B}_{j}{\bf P}_{j,1_S}(s-1)+{\bf C}_{j-1}(1)\Big(\delta_{j,1}{\bf P}_{0,1_S}(s)+(1-\delta_{j,1}){\bf h}_{j-1,1_S}(s)\Big)\Big),
    \\
    {\bf H}_{j,1_S} &=& {\bf I}_{N-j}-(1-\delta_{j,N-1}){\bf C}_{j}(0)-(1-\delta_{j,1}){\bf C}_{j-1}(1){\bf H}^{-1}_{j-1,1_S}{\bf C}_{j}(2),
\end{eqnarray*}
where ${\bf P}_{0,1_S}(s)$ is given by (\ref{eq:3}).}
\\
\par A point worth mentioning is that the structured form of (\ref{eq:4}) also allows us to derive an iterative scheme for computing the moments of the random number ${\cal R}_{exact,0}^S$ on the sample paths of process ${\cal X}$ satisfying $\{ {\cal R}_{exact,0}^R=0\}$. More particularly, in terms of the generating functions
\begin{eqnarray*}
    \varphi_{j,1_S}(x) &=& \sum_{s=0}^{\infty}x^s{\bf P}_{j,1_S}(s),\quad |x|\leq 1,
\end{eqnarray*}
for integers $j\in\{0,...,N-1\}$, Eqs. (\ref{eq:3})-(\ref{eq:4}) are solved to yield the expressions
\begin{eqnarray*}
    \varphi_{0,1_S}(x) &=& \left({\bf I}_N-{\bf C}_{0}(0)-x{\bf B}_{0}\right)^{-1}{\bf p}_{0,1_S},
    \\
    \varphi_{j,1_S}(x) &=& {\bf g}_{j,1_S}(x)+(1-\delta_{j,N-1}){\bf G}^{-1}_{j,1_S}(x){\bf C}_{j+1}(2)\varphi_{j+1,1_S}(x),
\end{eqnarray*}
for $j\in\{1,...,N-1\}$ and $|x|\leq 1$, where
\begin{eqnarray*}
    {\bf g}_{j,1_S}(x) &=& {\bf G}^{-1}_{j,1_S}(x)\Big({\bf p}_{j,1_S}  +{\bf C}_{j-1}(1)\Big(\delta_{j,1}\varphi_{0,1_S}(x)+(1-\delta_{j,1}){\bf g}_{j-1,1_S}(x)\Big)\Big),
    \\
    {\bf G}_{j,1_S}(x) &=& {\bf I}_{N-j}-x{\bf B}_{j}-(1-\delta_{j,N-1}){\bf C}_{j}(0)-(1-\delta_{j,1}){\bf C}_{j-1}(1){\bf G}^{-1}_{j-1,1_S}(x){\bf C}_{j}(2).
\end{eqnarray*}
\par In terms of the column vectors
\begin{eqnarray*}
    \varphi^{(n)}_{j,1_S} &=& \left.\frac{d^n\varphi_{j,1_S}(x)}{d x^n}\right|_{x=1},
\end{eqnarray*}
for $j\in\{0,...,N-1\}$, the computation of factorial moments of ${\cal R}_{exact,0}^S$ on the sample paths of process ${\cal X}$ satisfying $\{ {\cal R}_{exact,0}^R=0\}$ is possible, as shown in the result below.
\\ \\
{\bf Corollary 1} {\it For $n\in\mathbb{N}$, the column vectors $\{ \varphi^{(n)}_{j,1_S}: j\in\{0,...,N-1\} \}$ can be iteratively computed, starting with $\varphi^{(0)}_{j,1_S}={\bf 1}_N$, from the equalities
\begin{eqnarray*}
    \varphi^{(n)}_{0,1_S} &=& n\left( {\bf I}_N-{\bf C}_{0}(0)-{\bf B}_{0}\right)^{-1}{\bf B}_{0}\varphi^{(n-1)}_{0,1_S},
    \\
    \varphi^{(n)}_{j,1_S} &=& {\bf g}^{(n)}_{j,1_S}+(1-\delta_{j,N-1}){\bf G}^{-1}_{j,1_S}(1){\bf C}_{j+1}(2)\varphi^{(n)}_{j+1,1_S},
\end{eqnarray*}
for $j\in\{1,...,N-1\}$, where
\begin{eqnarray*}
    {\bf g}^{(n)}_{j,1_S} &=& {\bf G}^{-1}_{j,1_S}(1)\left( n{\bf B}_{j}\varphi^{(n-1)}_{j,1_S}+{\bf C}_{j-1}(1)\left(\delta_{j,1}\varphi^{(n)}_{0,1_S}+(1-\delta_{j,1}){\bf g}^{(n)}_{j-1,1_S}\right)\right),
\end{eqnarray*}
and ${\bf 1}_a$ is a column vector of order $a$ of 1's.}
\\
\par In particular, the first entry of the column vector
\begin{eqnarray*}
 \left({\bf I}_N-{\bf C}_{0}(0)-{\bf B}_{0}\right)^{-1}{\bf B}_{0}{\bf 1}_N
\end{eqnarray*}
is found to be the mean value of ${\cal R}_{exact,0}^S$ at an invasion time when the initially infected patient is colonized with sensitive bacteria; i.e., it corresponds to $E[ {\cal R}_{exact,0}^S | (S(0)+R(0),R(0))=(1,0)]$. This is derived by noting that, since $P_{(1,0),1_S}(s,r)=0$ if $r\in\mathbb{N}$, this mean value amounts to the expectation $E[ {\cal R}_{exact,0}^S 1\{ {\cal R}_{exact,0}^R=0\} | (S(0)+R(0),R(0))=(1,0), B(0)=1_S]$, and $\varphi^{(0)}_{0,1_S}=\varphi_{0,1_S}(1)$.
\subsection{The special case $B(0)=1_R$}
\label{subsec:2}
In order to evaluate the conditional probabilities $\{P_{(i,j),1_R}(s,r): i\in\{1,...,N\},$ $j\in\{1,...,i\} \}$, for integers $s,r\in\mathbb{N}_0$, and related moments of $({\cal R}_{exact,0}^S,{\cal R}_{exact,0}^R)$ we first introduce the column vectors ${\bf P}_{j,1_R}(s,r)$ and ${\bf p}_{j,1_R}$, for
$j\in\{1,...,N\}$, with $i$-th entries $P_{(j-1+i,j),1_R}(s,r)$ and $q^{-1}_{(j-1+i,j)}(\gamma+\tau_2+(1-m)\mu)$, respectively, if $i\in\{1,...,N-j+1\}$. We also consider
\begin{eqnarray*}
    \varphi^{(n)}_{j,1_R} &=& \left.\frac{d^n\varphi_{j,1_R}(x)}{d x^n}\right|_{x=1},
    \\
    \phi^{(n,\cdot)}_{j,1_R} &=& \left.\frac{\partial^n\phi_{j,1_R}(x,1)}{\partial x^n}\right|_{x=1},
    \\
    \phi^{(\cdot,m)}_{j,1_R} &=& \left.\frac{\partial^m\phi_{j,1_R}(1,y)}{\partial y^m}\right|_{y=1},
\end{eqnarray*}
for $j\in\{1,...,N\}$ and $n,m\in\mathbb{N}_0$, where $\varphi_{j,1_R}(x)=\sum_{s=0}^{\infty}x^s{\bf P}_{j,1_R}(s,0)$ and $\phi_{j,1_R}(x,y)=\sum_{s=0}^{\infty}\sum_{r=1}^{\infty}x^sy^r{\bf P}_{j,1_R}(s,r)$, for $|x|, |y|\leq 1$. We then use first-step analysis to yield the theorems below. The proofs mostly follow the argument yielding Theorem 1 and thus are omitted.
\\ \\
{\bf Theorem 2} {\it For $s\in\mathbb{N}_0$, the column vectors $\{{\bf P}_{j,1_R}(s,0): j\in\{1,...,N\} \}$ can be written in the form
\begin{eqnarray*}
    {\bf P}_{1,1_R}(s,0) &=& \left({\bf I}_N-{\bf D}_{1}(0)\right)^{-1}
    \Big(\delta_{s,0}{\bf p}_{1,1_R}+(1-\delta_{s,0}){\bf E}_{1}{\bf P}_{0,1_S}(s-1)\Big),
    \\
    {\bf P}_{j,1_R}(s,0) &=& {\bf h}_{j,1_R}(s,0)+(1-\delta_{j,N}){\bf H}^{-1}_{j,1_R}{\bf D}_{j+1}(2){\bf P}_{j+1,1_R}(s,0),
\end{eqnarray*}
for $j\in\{2,...,N\}$, where ${\bf P}_{0,1_S}(s-1)$ is evaluated from (\ref{eq:3}),
{\footnotesize
\begin{eqnarray*}
    {\bf h}_{j,1_R}(s,0) &=& {\bf H}^{-1}_{j,1_R}\Big(\delta_{s,0}{\bf p}_{j,1_R}+(1-\delta_{s,0}){\bf E}_{j}{\bf P}_{j-1,1_S}(s-1)+{\bf D}_{j-1}(1)\Big(\delta_{j,2}{\bf P}_{1,1_R}(s,0)+(1-\delta_{j,2}){\bf h}_{j-1,1_R}(s,0)\Big)\Big),
    \\
    {\bf H}_{j,1_R} &=& {\bf I}_{N-j+1}-(1-\delta_{j,N}){\bf D}_{j}(0)-(1-\delta_{j,2}){\bf D}_{j-1}(1){\bf H}^{-1}_{j-1,1_R}{\bf D}_{j}(2),
\end{eqnarray*}
}
and ${\bf D}_{j}(0)$, ${\bf D}_{j-1}(1)$, ${\bf D}_{j+1}(2)$ and ${\bf E}_{j}$ are suitably defined matrices of coefficients; see Appendix \ref{app:C}.}
\\ \\
{\bf Theorem 3} {\it For $s\in\mathbb{N}_0$ and $r\in\mathbb{N}$, the column vectors in $\{{\bf P}_{j,1_R}(s,r): j\in\{1,...,N\}\}$ have the form
\begin{eqnarray*}
    {\bf P}_{1,1_R}(s,r) &=& \left({\bf I}_N-{\bf D}_{1}(0)\right)^{-1}{\bf F}_{1}{\bf P}_{2,1_R}(s,r-1),
    \\
    {\bf P}_{j,1_R}(s,r) &=& {\bf h}_{j,1_R}(s,r)+(1-\delta_{j,N}){\bf H}^{-1}_{j,1_R}{\bf D}_{j+1}(2){\bf P}_{j+1,1_R}(s,r),
\end{eqnarray*}
for $j\in\{2,...,N\}$, where
\begin{eqnarray*}
    {\bf h}_{j,1_R}(s,r) &=& {\bf H}^{-1}_{j,1_R}\Big((1-\delta_{j,N}){\bf F}_{j}{\bf P}_{j+1,1_R}(s,r-1)\\
    &&\quad+{\bf D}_{j-1}(1)\Big(\delta_{j,2}(1-\delta_{j,N}){\bf P}_{1,1_R}(s,r)+(1-\delta_{j,2}){\bf h}_{j-1,1_R}(s,r)\Big)\Big),
\end{eqnarray*}
and matrices ${\bf F}_{j}$ are specified in Appendix \ref{app:C}.}
\\
\par The factorial conditional moments of ${\cal R}_{exact,0}^S$ on the sets $\{{\cal R}_{exact,0}^R=0\}$ and $\{{\cal R}_{exact,0}^R>0\}$, provided that $(S(0)+R(0),R(0))=(i,j)$ and $B(0)=1_R$, for integers $i\in\{j,...,N\}$ and $j\in\{1,...,N\}$, are given by the entries of $\varphi^{(n)}_{j,1_R}$ and $\phi^{(n,\cdot)}_{j,1_R}$, respectively. As a result, the mean value $E[{\cal R}_{exact,0}^S | (S(0)+R(0),R(0))=(1,1)]$ is given by the first entry of $\varphi^{(1)}_{1,1_R}+\phi^{(1,\cdot)}_{1,1_R}$.
\\ \\
{\bf Corollary 2} {\it For $n\in\mathbb{N}$, it is seen that
\begin{itemize}
    \item[{(i)}] The column vectors $\{\varphi^{(n)}_{j,1_R}: j\in\{1,...,N\} \}$ are found to satisfy
        \begin{eqnarray*}
            \varphi^{(n)}_{1,1_R} &=& \left({\bf I}_N-{\bf D}_{1}(0)\right)^{-1}{\bf E}_{1}\Big(\varphi^{(n)}_{0,1_S}+n \varphi^{(n-1)}_{0,1_S}\Big),
            \\
            \varphi^{(n)}_{j,1_R} &=& {\bf g}^{(n)}_{j,1_R}+(1-\delta_{j,N}){\bf H}^{-1}_{j,1_R}{\bf D}_{j+1}(2)\varphi^{(n)}_{j+1,1_R},
        \end{eqnarray*}
        for $j\in\{2,...,N\}$, where vectors $\varphi^{(n)}_{0,1_S}$, for $n\in\mathbb{N}_0$, are evaluated from Corollary 1 and
        \begin{eqnarray*}
            {\bf g}^{(n)}_{j,1_R} &=& {\bf H}^{-1}_{j,1_R}\Big({\bf E}_{j}\Big(n\varphi^{(n-1)}_{j-1,1_S}+\varphi^{(n)}_{j-1,1_S}\Big)+{\bf D}_{j-1}(1)\Big(\delta_{j,2} \varphi^{(n)}_{1,1_R} +(1-\delta_{j,2}){\bf g}^{(n)}_{j-1,1_R}\Big)\Big).
        \end{eqnarray*}
    \item[{(ii)}] The column vectors $\{\phi^{(n,\cdot)}_{j,1_R}: j\in\{1,...,N\} \}$ have the form
        \begin{eqnarray*}
            \phi^{(n,\cdot)}_{j,1_R} &=& {\bf l}^{(n,\cdot)}_{j,1_R}    + (1-\delta_{j,N}){\bf L}^{-1}_{j,1_R}\Big( {\bf F}_{j}+(1-\delta_{j,1}){\bf D}_{j+1}(2) \Big)\phi^{(n,\cdot)}_{j+1,1_R},
        \end{eqnarray*}
        where
        \begin{eqnarray*}
            {\bf l}^{(n,\cdot)}_{j,1_R} &=& {\bf L}^{-1}_{j,1_R}
            \left( (1-\delta_{j,1}){\bf D}_{j-1}(1){\bf l}^{(n,\cdot)}_{j-1,1_R} + (1-\delta_{j,N}){\bf F}_{j}\varphi^{(n)}_{j+1,1_R} \right),
        \end{eqnarray*}
        and matrices ${\bf L}_{j,1_R}$, for $j\in\{1,...,N\}$, are defined by
        \begin{eqnarray*}
            {\bf L}_{j,1_R} &=& {\bf I}_{N-j+1}-(1-\delta_{j,N}){\bf D}_{j}(0)-(1-\delta_{j,1}){\bf L}^{-1}_{j-1,1_R}{\bf D}_{j-1}(1) \Big( {\bf F}_{j-1}+(1-\delta_{j,2}){\bf D}_{j}(2)\Big).
        \end{eqnarray*}
    \end{itemize}}
\par Finally, the mean value $E[{\cal R}^R_{exact,0} | (S(0)+R(0),R(0))=(1,1)]$ can be derived as the first entry of $\phi^{(\cdot,1)}_{1,1_R}$ and, in a more general way, column vectors $\phi^{(\cdot,m)}_{j,1_R}$, for $j\in\{1,...,N\}$, record the $m$-th factorial conditional moments of ${\cal R}_{exact,0}^R$, provided that $(S(0)+R(0),R(0))=(i,j)$ and $B(0)=1_R$, for $i\in\{j,...,N\}$.
\\ \\
{\bf Corollary 3} {\it For $m\in\mathbb{N}$, the column vectors $\{\phi^{(\cdot,m)}_{j,1_R}: j\in\{1,...,N\} \}$ are specified by
\begin{eqnarray*}
    \phi^{(\cdot,m)}_{j,1_R} &=& {\bf l}^{(\cdot,m)}_{j,1_R}
    +(1-\delta_{j,N}){\bf L}^{-1}_{j,1_R}\Big({\bf F}_{j}+(1-\delta_{j,1}){\bf D}_{j+1}(2)\Big)\phi^{(\cdot,m)}_{j+1,1_R},
\end{eqnarray*}
where
\begin{eqnarray*}
    {\bf l}^{(\cdot,m)}_{j,1_R} &=& {\bf L}^{-1}_{j,1_R}\left( (1-\delta_{j,1}){\bf D}_{j-1}(1){\bf l}^{(\cdot,m)}_{j-1,1_R}+(1-\delta_{j,N}){\bf F}_{j} \left(\delta_{m,1}\varphi_{j+1,1_R}(1)+m\phi^{(\cdot,m-1)}_{j+1,1_R}\right) \right).
\end{eqnarray*}}
\section{Numerical experiments and discussion}
\label{sec:4}
\par In this section, we present some numerical experiments to illustrate the variability of the probability law of the random contributions ${\cal R}_{exact,0}^R$ and ${\cal R}_{exact,0}^S$ to the exact reproduction number ${\cal R}_{exact,0}$, mainly as a function of the proportion $m$ of individuals who enter the hospital carrying sensitive bacteria. We consider a hospital ward that consists of one marked bed, where a patient colonized with resistant bacteria is accommodated at time $t=0$, and nineteen beds accommodating initially uncolonized patients; i.e., $R(0)=1$, $S(0)=0$ and $X(0)=19$. It is assumed that inpatients contact each other, on average, in $1$ \emph{day}, whence the per capita infection rate is given by $\beta=N^{-1}$ with $N=20$.
\begin{figure}
	\centering
	\begin{center}
        \includegraphics[width=0.8\textwidth]{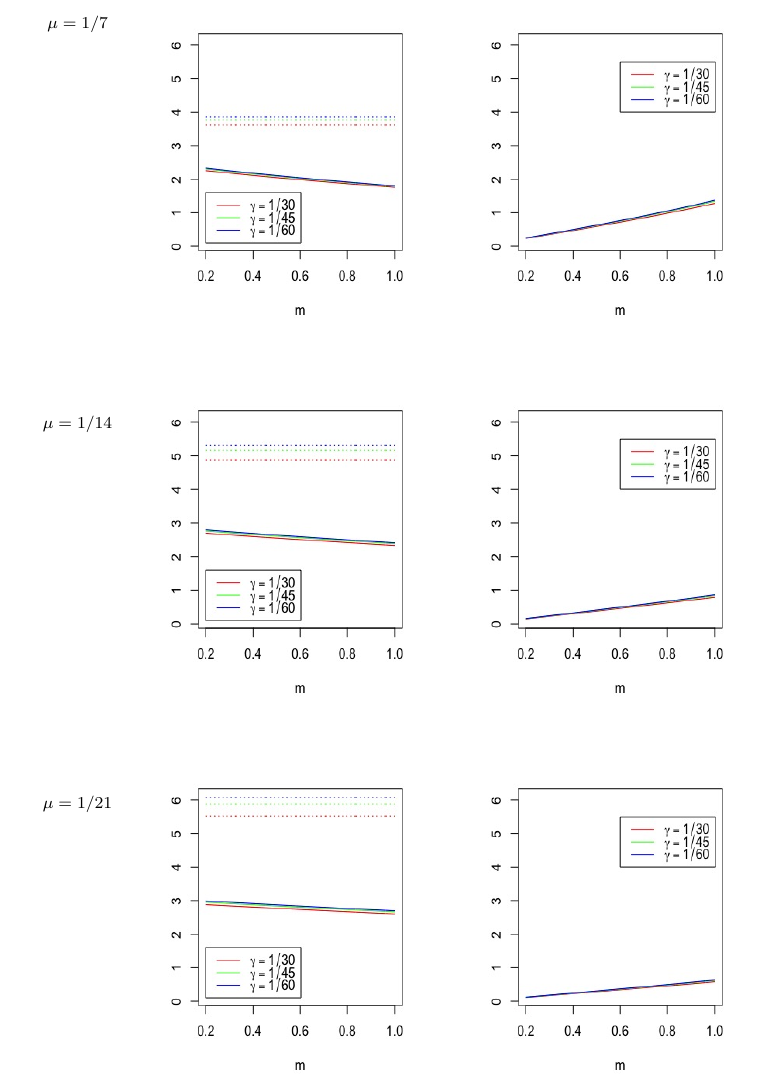}
	\end{center}
	\caption{The basic reproduction number $R_0$ (left column, dotted lines), and expected values $E[{\cal R}_{exact,0}^R | (S(0)+R(0),R(0))=(1,1)]$ (left column, solid lines) and $E[{\cal R}_{exact,0}^S | (S(0)+R(0),R(0))=(1,1)]$ (right column) as a function of $m$, for values of $\mu^{-1}\in\{7, 14, 21\}$ \emph{days}, $\gamma^{-1}\in\{30, 45, 60\}$ \emph{days}, $\tau_1^{-1}=5$ \emph{days}, $\tau_2^{-1}=10$ \emph{days} and $c=0.05$.}
	\label{fig:2}
\end{figure}
\begin{figure}
	\centering
	\begin{center}
        \includegraphics[width=0.8\textwidth]{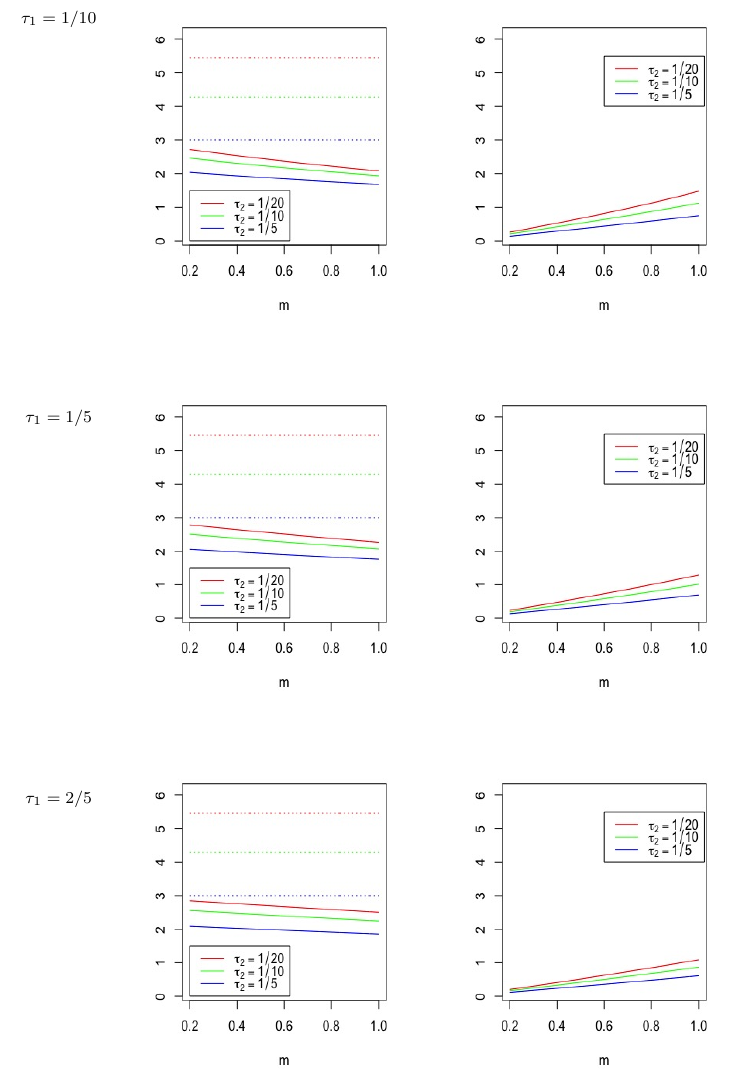}
	\end{center}
	\caption{The basic reproduction number $R_0$ (left column, dotted lines), and expected values $E[{\cal R}_{exact,0}^R | (S(0)+R(0),R(0))=(1,1)]$ (left column, solid lines) and $E[{\cal R}_{exact,0}^S | (S(0)+R(0),R(0))=(1,1)]$ (right column) as a function of $m$, for values of $\tau_1^{-1}\in\{2.5, 5, 10\}$ \emph{days}, $\tau_2^{-1}\in\{5, 10, 20\}$ \emph{days}, $\mu^{-1}=14$ \emph{days}, $\gamma^{-1}=30$ \emph{days} and $c=0.05$.}
	\label{fig:3}
\end{figure}
\par
\begin{figure}
	\centering
	\begin{center}
        \includegraphics[width=0.8\textwidth]{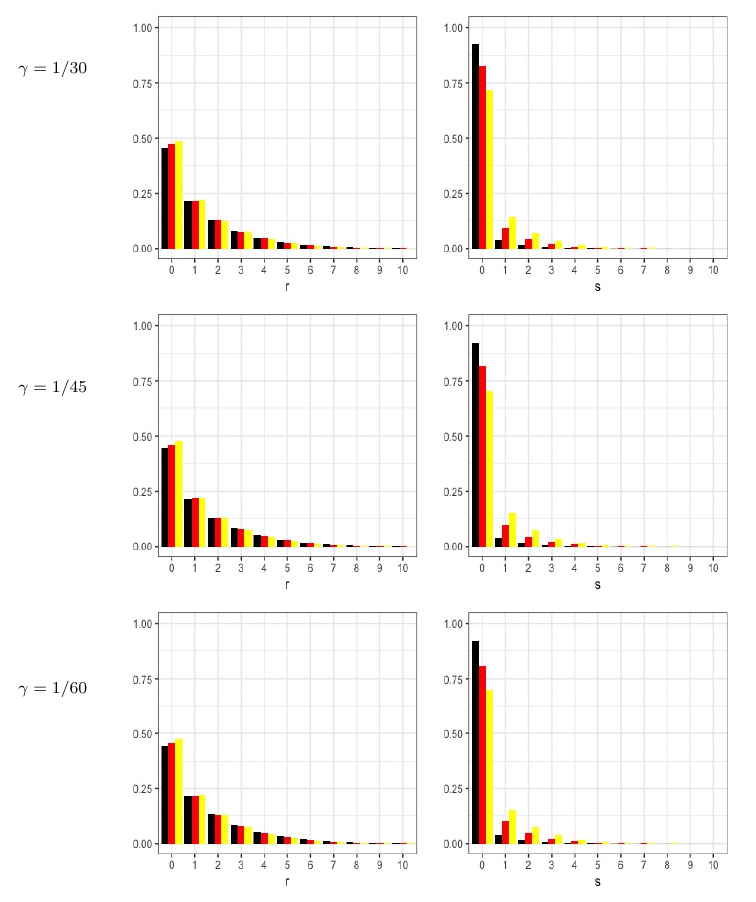}
	\end{center}
	\caption{The mass functions $\{ P({\cal R}_{exact,0}^R = r | (S(0)+R(0),R(0))=(1,1)) : r\in\mathbf{N}_0\}$ (left column) and $\{ P({\cal R}_{exact,0}^S = s | (S(0)+R(0),R(0))=(1,1)) : s\in\mathbf{N}_0\}$ (right column) in scenarios with proportion $m=0.2$ (black), $0.5$ (red) and $0.8$ (yellow), for values of $\mu^{-1}=14$ \emph{days}, $\gamma^{-1}\in\{30, 45, 60\}$ \emph{days}, $\tau_1^{-1}=5$ \emph{days}, $\tau_2^{-1}=10$ \emph{days} and $c=0.05$.}
	\label{fig:4}
\end{figure}

\begin{figure}
	\centering
	\begin{center}
        \includegraphics[width=0.8\textwidth]{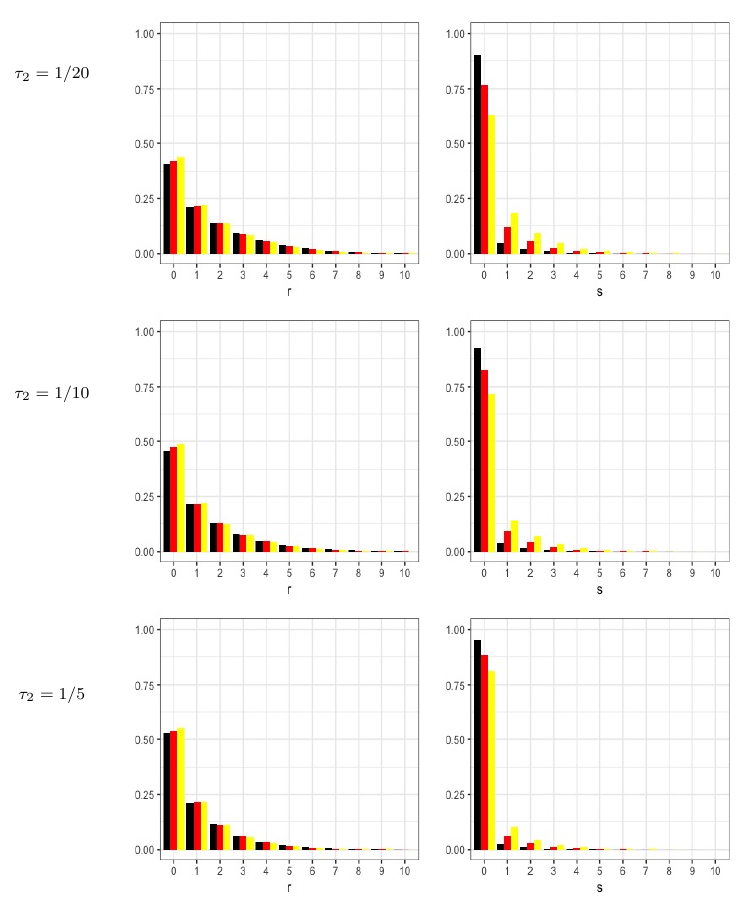}
	\end{center}
	\caption{The mass functions $\{ P({\cal R}_{exact,0}^R = r | (S(0)+R(0),R(0))=(1,1)) : r\in\mathbf{N}_0\}$ (left column) and $\{ P({\cal R}_{exact,0}^S = s | (S(0)+R(0),R(0))=(1,1)) : s\in\mathbf{N}_0\}$ (right column) in scenarios with proportion $m=0.2$ (black), $0.5$ (red) and $0.8$ (yellow), for values of $\tau_1^{-1}=5$ \emph{days}, $\tau_2^{-1}\in\{5, 10, 20\}$ \emph{days}, $\mu^{-1}=14$ \emph{days}, $\gamma^{-1}=30$ \emph{days} and $c=0.05$.}
	\label{fig:5}
\end{figure}

 In our first numerical experiments, the fitness difference between sensitive and resistant bacterial strains is relatively small ($c=0.05$), and scenarios in Figures \ref{fig:2}-\ref{fig:5} are specified from suitable choices of the average duration of hospital stay $\mu^{-1}\in\{7, 14, 21\}$ \emph{days}, the average time from admission or colonization until spontaneous clearance of bacterial carriage $\gamma^{-1}\in\{30, 45, 60\}$ \emph{days}, and the proportion $m\in [0.2,1.0]$ of admitted already colonized with sensitive bacteria. Drugs 1 and 2 are assumed to be effective on colonized patients, on average, in $\tau_1^{-1}\in\{2.5, 5, 10\}$ \emph{days} and $\tau_2^{-1}\in\{5, 10, 20\}$ \emph{days}, respectively. For details on these values of parameters and published sources for them, we refer the reader to Ref. \cite{Lipsitch2000} and references therein.
\par Figure \ref{fig:2} (respectively, Figure \ref{fig:3}) illustrates the dynamics of colonization with resistant and sensitive bacteria in terms of the basic reproduction number $R_0$, and of the expectations of the random numbers ${\cal R}_{exact,0}^R$ and ${\cal R}_{exact,0}^S$, which are plotted as a function of the proportion $m$, for values of $\mu^{-1}\in\{7, 14, 21\}$ \emph{days}, $\gamma^{-1}\in\{30, 45, 60\}$ \emph{days}, $\tau_1^{-1}=5$ \emph{days} and $\tau_2^{-1}=10$ \emph{days} (respectively, $\tau_1^{-1}\in\{2.5, 5, 10\}$ \emph{days}, $\tau_2^{-1}\in\{5, 10, 20\}$ \emph{days}, $\mu^{-1}=14$ \emph{days} and $\gamma^{-1}=30$ \emph{days}). A first important observation is that, in our experiments, values of $R_0$ are found to be greater than the corresponding values for the expectation of ${\cal R}_{exact,0}^R$, regardless of the parameters. This observation is counterintuitive because the random length of the interval during which secondary cases contribute to ${\cal R}_{exact,0}^{R}$ is likely to be longer than the length of that involved in $R_0$. However, it can be understood as a consequence of the fact that $R_0$ is formally intended to be an index of the potential, but not exact, contagiousness of the resistant bacteria at early stages of the epidemic and, therefore, its values in Ref. \cite{Lipsitch2000} are only affected by the dynamics of the inpatient-to-inpatient contact process until either the treatment with drug 2 or other non-therapeutic reasons clear carriage of resistant bacteria on the initially colonized patient, or this patient's hospital stay ends.
\par Figures \ref{fig:2}-\ref{fig:3} illustrate the impact of $m$, the proportion of patients admitted to the hospital who are already colonized with sensitive bacteria, on the transmission dynamics. In particular, the expected value of the exact reproduction number corresponding to the resistant strain, ${\cal R}_{exact,0}^R$, decreases with increasing values of $m$, which is directly related to the assumption of cross-immunity, and highlights the competition dynamics between both bacterial strains. That is, the more patients colonized with sensitive bacteria that arrive into the hospital, the more difficult is for the resistant bacteria to propagate in the hospital ward. Interestingly, for a relatively small efficacy of drug 1 (i.e., $\tau_1=1/10$ in Figure \ref{fig:3}), this impact is less significant under increasing efficacies of drug 2 (i.e., increasing $\tau_2$ in Figure \ref{fig:3}). We note that increasing values of $\tau_2$ would ultimately lead to $\tau_2\approx \tau_1+\tau_2$, making both strains \emph{equally competitive}, and thus the impact of $m$ is less significant in these scenarios. Similar (but opposite) behaviours can be observed in these figures for ${\cal R}_{exact,0}^S$, for the same reasons. In a similar way, an intervention based on reducing the hospital stay will reduce the carriage of resistant bacteria, while the prevalence of the sensitive strain will become more significant. A more frequent spontaneous clearance of bacterial carriage will also decrease the prevalence of both resistant and sensitive strains, although no very significant changes are observed in our numerical experiments in terms of expectations of ${\cal R}_{exact,0}^R$ and ${\cal R}_{exact,0}^S$.
\par Not surprisingly, the use of more effective drugs for which there is no resistance (i.e., increasing values of $\tau_2$) will result in a reduction in the prevalence of both resistant and sensitive bacterial strains. More surprising, on at least first consideration, is the prediction that, at an early stage of the epidemic, the effectiveness of drug 1 will not necessarily result in a remarkable variation in the prevalence of resistant bacteria. More concretely, the model predicts that the use of less effective drugs for which there is resistance will lead to a more significant reduction in the prevalence of resistant bacteria only when more individuals already colonized with sensitive bacteria are admitted to the hospital.
\begin{figure}
	\centering
	\begin{center}
        \includegraphics[width=0.8\textwidth]{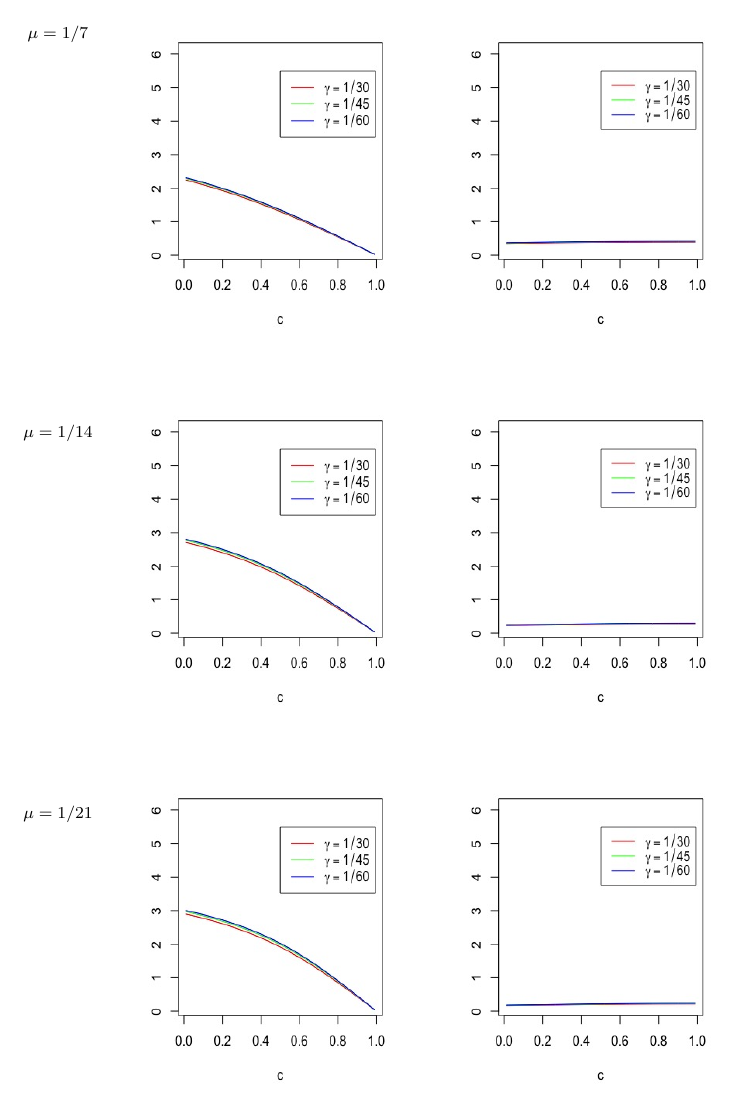}
	\end{center}
	\caption{The expected values $E[{\cal R}_{exact,0}^R | (S(0)+R(0),R(0))=(1,1)]$ (left column) and $E[{\cal R}_{exact,0}^S | (S(0)+R(0),R(0))=(1,1)]$ (right column) as a function of $c$, for values of $\mu^{-1}\in\{7, 14, 21\}$ \emph{days}, $\gamma^{-1}\in\{30, 45, 60\}$ \emph{days}, $\tau_1^{-1}=5$ \emph{days}, $\tau_2^{-1}=10$ \emph{days} and $m=0.3$.}
	\label{fig:6}
\end{figure}
\begin{figure}
	\centering
	\begin{center}
        \includegraphics[width=0.8\textwidth]{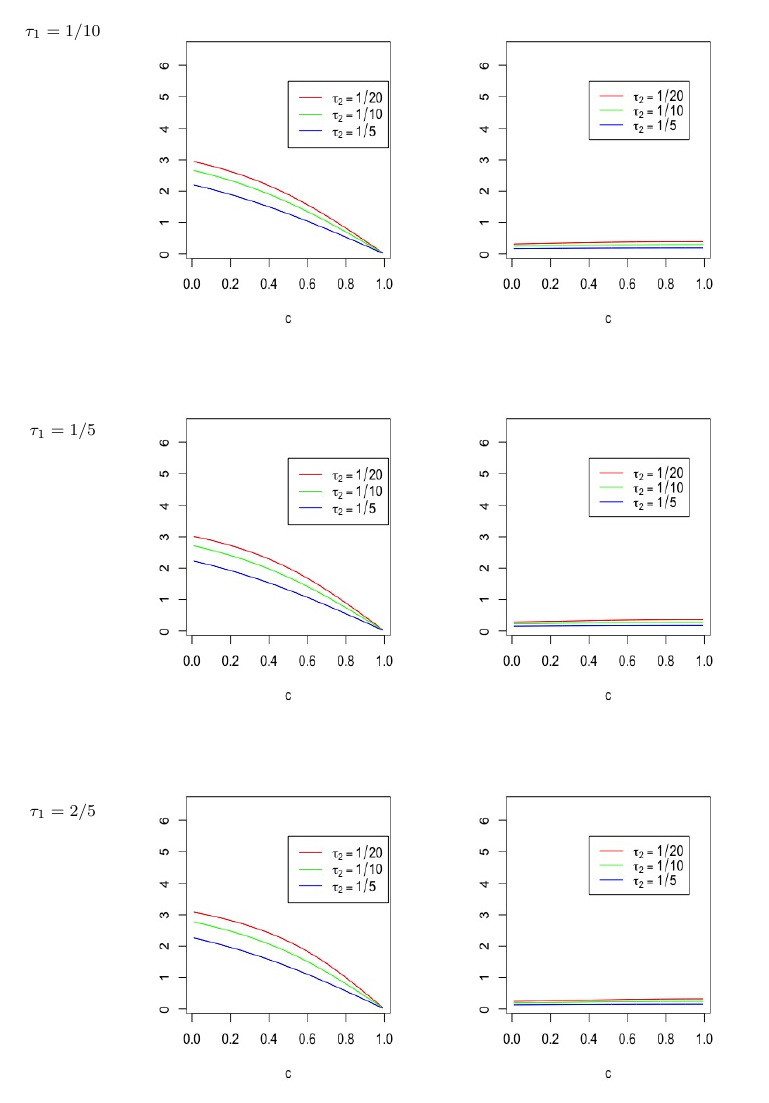}
	\end{center}
	\caption{The expected values $E[{\cal R}_{exact,0}^R | (S(0)+R(0),R(0))=(1,1)]$ (left column) and $E[{\cal R}_{exact,0}^S | (S(0)+R(0),R(0))=(1,1)]$ (right column) as a function of $c$, for values of $\tau_1^{-1}\in\{2.5, 5, 10\}$ \emph{days}, $\tau_2^{-1}\in\{5, 10, 20\}$ \emph{days}, $\mu^{-1}=14$ \emph{days}, $\gamma^{-1}=30$ \emph{days} and $m=0.3$.}
	\label{fig:7}
\end{figure}
\par A more detailed description of ${\cal R}_{exact,0}^R$ and of ${\cal R}_{exact,0}^S$ is displayed in Figure \ref{fig:4} (respectively, Figure \ref{fig:5}) in terms of mass functions, instead of expected values, in selected scenarios with $m\in\{0.2, 0.5, 0.8\}$, for values of $\mu^{-1}=14$ \emph{days}, $\gamma^{-1}\in\{30, 45, 60\}$ \emph{days}, $\tau_1^{-1}=5$ \emph{days} and $\tau_2^{-1}=10$ \emph{days} (respectively, $\tau_1^{-1}=5$ \emph{days}, $\tau_2^{-1}\in\{5, 10, 20\}$ \emph{days}, $\mu^{-1}=14$ \emph{days} and $\gamma^{-1}=30$ \emph{days}). The random numbers ${\cal R}_{exact,0}^R$ and ${\cal R}_{exact,0}^S$ are seen to have unimodal distributions with a clear peak at points $r=0$ and $s=0$, meaning that events $\{{\cal R}_{exact,0}^R=0\}$ and $\{{\cal R}_{exact,0}^S=0\}$ occur more frequently than the others. This does not mean that the corresponding expectations of ${\cal R}_{exact,0}^R$ and ${\cal R}_{exact,0}^S$ are necessarily less than one, as shown in Figures \ref{fig:2} and \ref{fig:3}. In our experiments, the tail of the distribution of ${\cal R}_{exact,0}^R$ is heavier than that of ${\cal R}_{exact,0}^S$, showing that the spread of the resistant strain appears to be more likely to occur than the spread of the sensitive one in an early stage of the epidemic. Despite this, the model predicts that ${\cal R}_{exact,0}^R$ will also take small values with high probability, just as ${\cal R}_{exact,0}^S$ will take large values with relatively significant, but small, probability.
\begin{figure}
	\centering
	\begin{center}
        \includegraphics[width=0.8\textwidth]{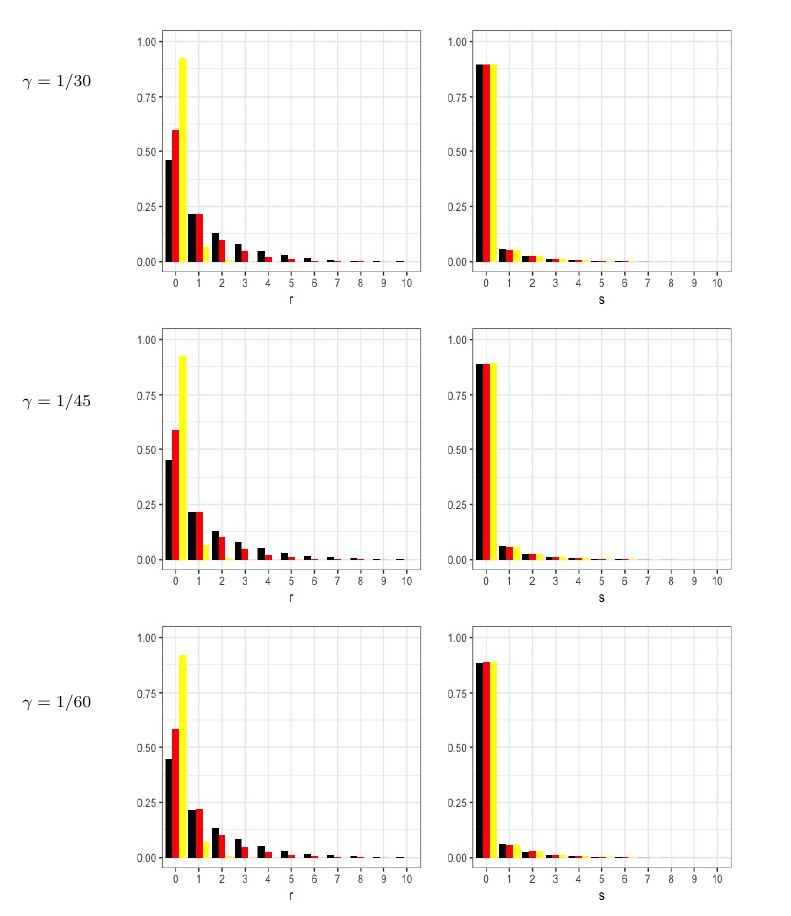}
	\end{center}
	\caption{The mass functions $\{ P({\cal R}_{exact,0}^R = r | (S(0)+R(0),R(0))=(1,1)) : r\in\mathbf{N}_0\}$ (left column) and $\{ P({\cal R}_{exact,0}^S = s | (S(0)+R(0),R(0))=(1,1)) : s\in\mathbf{N}_0\}$ (right column) in scenarios with fitness cost $c=0.05$ (black), $0.5$ (red) and $0.95$ (yellow), for values of $\mu^{-1}=14$ \emph{days}, $\gamma^{-1}\in\{30, 45, 60\}$ \emph{days}, $\tau_1^{-1}=5$ \emph{days}, $\tau_2^{-1}=10$ \emph{days} and $m=0.3$.}
	\label{fig:8}
\end{figure}
\begin{figure}
	\centering
	\begin{center}
        \includegraphics[width=0.8\textwidth]{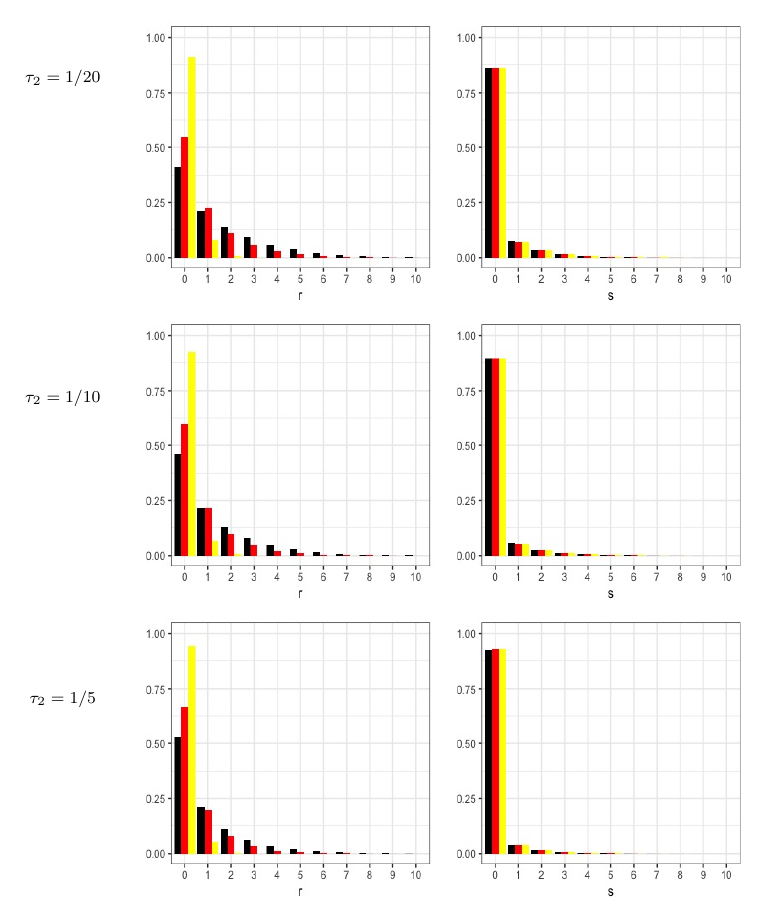}
	\end{center}
	\caption{The mass functions $\{ P({\cal R}_{exact,0}^R = r | (S(0)+R(0),R(0))=(1,1)) : r\in\mathbf{N}_0\}$ (left column) and $\{ P({\cal R}_{exact,0}^S = s | (S(0)+R(0),R(0))=(1,1)) : s\in\mathbf{N}_0\}$ (right column) in scenarios with fitness cost $c=0.05$ (black), $0.5$ (red) and $0.95$ (yellow), for values of $\tau_1^{-1}=5$ \emph{days}, $\tau_2^{-1}\in\{5, 10, 20\}$ \emph{days}, $\mu^{-1}=14$ \emph{days}, $\gamma^{-1}=30$ \emph{days} and $m=0.3$.}
	\label{fig:9}
\end{figure}
\par In Figures \ref{fig:6}-\ref{fig:9}, the interest is in analysing the impact of the fitness cost, $c$, on the transmission dynamics.
In particular, the expected values of ${\cal R}_{exact,0}^R$ and ${\cal R}_{exact,0}^S$ are plotted in Figure \ref{fig:6} (respectively, Figure \ref{fig:7}) as a function of $c$, for values $\mu^{-1}\in\{7, 14, 21\}$ \emph{days}, $\gamma^{-1}\in\{30, 45, 60\}$ \emph{days}, $\tau_1^{-1}=5$ \emph{days}, $\tau_2^{-1}=10$ \emph{days} (respectively, $\tau_1^{-1}\in\{2.5, 5, 10\}$ \emph{days}, $\tau_2^{-1}\in\{5, 10, 20\}$ \emph{days}, $\mu^{-1}=14$ \emph{days}, $\gamma^{-1}=30$ \emph{days}) and $m=0.3$. Figures \ref{fig:6}-\ref{fig:7} show how increasing values of $c$ lead to less transmission of the resistant strain, with no transmission at all for $c=1.0$, as one would expect. Once again, the impact of the fitness cost is influenced by the efficacy of drugs 1 and 2. For $c=0.0$ (no transmission advantage for the sensitive strain), increasing values of $\tau_2$ lead to smaller expected values of ${\cal R}_{exact,0}^R$ and ${\cal R}_{exact,0}^S$ in Figure \ref{fig:7}. Moreover, the impact of $c$ on the behaviour of ${\cal R}_{exact,0}^S$ is much less significant, as one would expect, since $c$ represents the fitness cost of the resistant bacterial strain.
\par Figure \ref{fig:8} (respectively, Figure \ref{fig:9}) is related to the mass functions of ${\cal R}_{exact,0}^R$ and ${\cal R}_{exact,0}^S$, and scenarios with fitness cost $c\in\{0.05,0.5,0.95\}$, for values of $\mu^{-1}=14$ \emph{days}, $\gamma^{-1}\in\{30, 45, 60\}$ \emph{days}, $\tau_1^{-1}=5$ \emph{days}, $\tau_2^{-1}=10$ \emph{days} (respectively, $\tau_1^{-1}=5$ \emph{days}, $\tau_2^{-1}\in\{5, 10, 20\}$ \emph{days}, $\mu^{-1}=14$ \emph{days}, $\gamma^{-1}=30$ \emph{days}) and $m=0.3$. In a similar manner to Figures 6-7, the significant impact of $c$ on the probability distribution of ${\cal R}_{exact,0}^R$ can also be seen in our results in Figures \ref{fig:8}-\ref{fig:9}, where increasing fitness costs lead to distributions more concentrated around $r=0$ infections. This highlights how bacterial strains which are at an advantage due to being resistant to one of the drugs, but at a significant disadvantage due to a high fitness cost decreasing its transmission rate, would not be competitive to cause significant outbreaks. On the other hand, the impact of $c$ on the probability distribution of ${\cal R}_{exact,0}^S$ is negligible in Figures \ref{fig:8}-\ref{fig:9}.
\section{Conclusions}
\label{sec:5}
\par In this work, we have developed a Markov chain version of the deterministic model for the spread of antibiotic-resistant bacteria in hospital settings proposed in \cite{Lipsitch2000}. We have focused our analysis on the exact number of secondary infections caused by all patients using a marked bed, which is initially occupied by an infected patient, until this bed is eventually occupied by a susceptible one. This stochastic descriptor allows one to estimate the ``{\it infectivity of a bed}'' in the ward, rather than of a single patient, by taking into account that patients arriving into the hospital ward can already be colonized with certain probability, and infect others. Our approach allows one to split the reproduction number into two random variables, depending on the type of infections caused (i.e., by either the antibiotic-sensitive or resistant bacterial strains), and to compute the joint probability distribution of these.
\par Our numerical results highlight the competition dynamics expected between the antibiotic-sensitive and antibiotic-resistant bacterial strains, which is directly related to the assumption of cross-immunity. Our results also show that the probability distributions of these exact reproduction numbers can, in fact, be very wide. Interestingly, one can find parameter regimes where the mean exact reproduction number is less than one, but there is a non-negligible probability of the marked bed causing a significant number of infections in early times, and regimes where the mean exact reproduction number is greater than one, but there is a significant probability of the marked bed causing zero infections. This could have a direct impact on the probability of an outbreak happening from a particular bed, and highlights the need to analyse these stochastic descriptors as random variables (i.e., in terms of probability distributions) rather than mean quantities.


\section*{Acknowledgments}
This research was supported by the Government of Spain (Ministry of Science and Innovation), project PGC2018-097704-B-I00.
FACCC was partially funded by Funda\c{c}\~{a}o para a Ci\^{e}ncia e a Tecnologia (Portugal), projects UID/MAT/00297/2019, UIDB/00297/2020, UIDP/00297/2020 and 2022.03091.PTDC.


\section*{Conflict of interest}
The authors declare no potential conflict of interests.

\bibliography{wileyNJD-HARVARD}

\begin{thebibliography}{}
\bibitem{Allen03}
Allen LJS (2003)
An Introduction to Stochastic Processes with Applications to Biology.
Pearson Education, Inc., Upper Saddle River, NJ.

\bibitem{Almaraz19}
Almaraz E, G\'{o}mez-Corral A (2019)
Number of infections suffered by a local individual in a two-strain SIS model with partial cross-immunity.
Mathematical Methods in the Applied Sciences 42: 4318--4330.

\bibitem{Amador18}
Amador J, L\'{o}pez-Herrero MJ (2018)
Cumulative and maximum epidemic sizes for a nonlinear SEIR stochastic model with limited resources.
Discrete and Continuous Dynamical Systems--Series B 23: 3137--3151.

\bibitem{Amador19}
Amador J, Armesto D, G\'{o}mez-Corral A (2019)
Extreme values in SIR epidemic models with two strains and cross--immunity.
Mathematical Biosciences and Engineering 16: 1992--2022.

\bibitem{Amador20}
Amador J, G\'{o}mez-Corral A (2020)
A stochastic model with two quarantine states and a limited carrying capacity for quarantine.
Physica A 544: 121899.

\bibitem{Artalejo13}
Artalejo JR, L\'{o}pez-Herrero MJ (2013)
On the exact measure of disease spread in stochastic epidemic models.
Bulletin of Mathematical Biology 75: 1031--1050.

\bibitem{Bagkur22}
Bagkur C, Guler E, Kaymakamzade B, Hincal E, Suer K (2022)
Near future perspective of ESBL-producing {\it Klebsiella pneumoniae} strains using mathematical modeling.
Computer Modeling in Engineering \& Sciences 130: 111--132.

\bibitem{Breijyeh20}
Breijyeh Z, Jubeh B, Karaman R (2020)
Resistance of {\it Gram-negative} bacteria to current antibacterial agents and approaches to resolve it.
Molecules 25: 1340.

\bibitem{Calderwood23}
Calderwood M, Anderson D, Bratzler D, Dellinger E, Garcia-Houchins S, Maragakis L, Nyquist AC, Perkins KM, Preas MA, Saiman L, Schaffzin JK, Schweizer M, Yokoe DS, Kaye K (2023)
Strategies to prevent surgical site infections in acute-care hospitals: 2022 Update.
Infection Control \& Hospital Epidemiology  44: 695--720.

\bibitem{Chalub09}
Chalub FACC, Souza MO (2009)
From discrete to continuous evolution models: A unifying approach to drift-diffusion and replicator dynamics.
Theoretical Population Biology 76: 268--277.

\bibitem{Chalub11}
Chalub FACC, Souza, MO (2011)
The SIR epidemic model from a PDE point of view.
Mathematical and Computer Modelling 53: 1568--1574.

\bibitem{Chalub14}
Chalub FACC, Souza MO (2014)
Discrete and continuous SIS epidemic models: A unifying approach.
Ecological Complexity 18: 83--95.

\bibitem{Chalub14b}
Chalub FACC, Souza MO (2014)
The frequency-dependent Wright-Fisher model: Diffusive and non-diffusive approximations.
Journal of Mathematical Biology 68: 1089--1133.

\bibitem{Chugonova23}
Chugunova M, Taranets R, Vasylyeva N (2023)
Initial-boundary value problems for conservative Kimura-type equations: Solvability, asymptotic and conservation law.
Journal of Evolution Equations 23: 17.

\bibitem{D'Agata02}
D'Agata EMC, Horn MA, Webb GF (2002)
The impact of persistent gastrointestinal colonization on the transmission dynamics of vancomycin-resistant {\it Enterococci}.
Journal of Infectious Diseases 185: 766--773.

\bibitem{D'Agata07}
D'Agata EMC, Magal P, Olivier D, Ruan S, Webb GF (2007)
Modeling antibiotic resistance in hospitals: The impact of minimizing treatment duration.
Journal of Theoretical Biology 249: 487--499.

\bibitem{Danilkina18}
Danilkina O, Souza MO, Chalub FACC (2018)
Conservative parabolic problems: Nondegenerated theory and degenerated examples from population dynamics.
Mathematical Methods in the Applied Sciences 41: 4391--4406.

\bibitem{DeNitto96}
De Nitto Person\`{e} V, Grassi V (1996)
Solution of finite QBD processes.
Journal of Applied Probability 33: 1003--1010.

\bibitem{Fernandez21}
Fern\'{a}ndez-Peralta R, G\'{o}mez-Corral A (2021)
A structured Markov chain model to investigate the effects of pre-exposure vaccines in tuberculosis control.
Journal of Theoretical Biology 509: 110490.

\bibitem{Gamboa22}
Gamboa M,  L\'{o}pez-Herrero MJ (2022)
Measures to asses a warning vaccination level in a stochastic SIV model with imperfect vaccine.
Studies in Applied Mathematics 148: 1411-1438.

\bibitem{Gaver84}
Gaver DP, Jacobs PA, Latouche G (1984)
Finite birth-and-death models in randomly changing environments.
Advances in Applied Probability 16: 715--731.

\bibitem{GC18}
G\'{o}mez-Corral A, L\'{o}pez-Garc\'{\i}a M (2018)
Perturbation analysis in finite LD-QBD processes and applications to epidemic models.
Numerical Linear Algebra with Applications 25: e2160.

\bibitem{GC20}
G\'{o}mez-Corral A, L\'{o}pez-Garc\'{\i}a M, L\'{o}pez-Herrero MJ, Taipe D (2020)
On firt-passage times and sojourn times in finite QBD processes and their applications in epidemics.
Mathematics 8: 1718.

\bibitem{GC21}
G\'{o}mez-Corral A, L\'{o}pez-Garc\'{\i}a M, Rodr\'{\i}guez-Bernal MT (2021)
On time-discretized versions of SIS epidemic models: A comparative analysis.
Journal of Mathematical Biology 82: 46.

\bibitem{GC23}
G\'{o}mez-Corral A, Palacios-Rodr\'{\i}guez F, Rodr\'{\i}guez-Bernal MT (2022)
On the distribution of the exact reproduction number in SIS epidemic models with vertical transmission.
Under review.

\bibitem{Gygli17}
Gygli SM, Borrell S, Trauner A, Gagneux S (2017)
Antimicrobial resistance in {\it Mycobacterium tuberculosis}: mechanistic and evolutionary perspectives.
FEMS Microbiology Reviews 41: 354--373.

\bibitem{Haaber17}
Haaber J, Penades JR, Ingmer H (2017)
Transfer of antibiotic resistance in {\it Staphylococcus aureus}.
Trends in Microbiology 25: 327--337.

\bibitem{Lipsitch2000}
Lipsitch M, Bergstrom CT, Levin BR (2000)
The epidemiology of antibiotic resistance in hospitals: Paradoxes and prescriptions.
Proceedings of the National Academic of Sciences 976: 1938--1943.

\bibitem{Lipsitch09}
Lipsitch M, Colijn C, Cohen T, Hanage WP, Fraser C (2009)
No coexistence for free: neutral null models for multistrain pathogens.
Epidemics 1: 2--13.

\bibitem{Miller14}
Miller WR, Munita JM, Arias CA (2014)
Mechanisms of antibiotic resistance in {\it enterococci}.
Expert Review of Anti-infective Theraphy 12: 1221--1236.

\bibitem{Murray22}
Murray CJL, Ikuta KS, Sharara F, Swetschinski L, Robles Aguilar G et al. (2022)
Global burden of bacterial antimicrobial resistance in 2019: A systematic analysis.
The Lancet 399: 629--655,

\bibitem{Niewiadomska19}
Niewiadomska AM, Jayabalasingham B, Seidman JC, Willem L, Grenfell B, Spiro D, Voboud C (2019)
Population-level mathematical modeling of antimicrobial resistance: A systematic review.
BMC Medicine 17: 81.

\bibitem{Pak21}
Pak H, Maghsoudi LH, Ahmadinejad M, Kabir K, Soltanian A, Vasi M (2021)
Assessment of prophylactic Antibiotic Prescription Pattern in elective surgery patients in accordance with national and international guidelines.
International Journal of Surgery Open 29: 40--44.

\bibitem{Seigal2017}
Seigal A, Mira P, Sturmfels B, Barlow M (2017)
Does antibiotic resistance evolve in hospitals?
Bulletin of Mathematical Biology 79: 191--208.

\bibitem{Spicknall13}
Spicknall IH, Foxman B, Marrs CF, Eisenberg JNS (2013)
A modeling framework for the evolution and spread of antibiotiv resistance: Literature review and model categorization.
American Journal of Epidemiology 178: 508--520.

\bibitem{Techitnutsarut21}
Techitnutsarut P, Chamchod F (2021)
Modeling bacterial resistance to antibiotics: Bacterial conjugation and drug effects.
Advances in Difference Equations 2021: 290.

\bibitem{Webb05}
Webb GF, D'Agata EMC, Magal P, Ruan S (2005)
A model of antibiotic-resistant bacterial epidemics in hospitals.
Proceedings of the National Academic of Sciences 102: 13343--13348.
\end{thebibliography}


\appendix

\section{Expressions for matrices ${\bf B}_{0}$ and ${\bf C}_{0}(0)$ in Eq. (5)}
\label{app:A}
From Eqs. (\ref{eq:1})-(\ref{eq:2}) it is seen that ${\bf B}_{0}$ and ${\bf C}_{0}(0)$ are square matrices of order $N$ and have the form
\begin{eqnarray*}
    {\bf B}_{0} &=& \left(\begin{array}{ccccc}
        \frac{m\mu}{q_{(1,0)}} & \frac{(N-1)\beta}{q_{(1,0)}} &                        &        & \\
                               & \frac{m\mu}{q_{(2,0)}} & \frac{(N-2)\beta}{q_{(2,0)}} &        & \\
                               &                        & \ddots                       & \ddots & \\
                               &                        &                              & \frac{m\mu}{q_{(N-1,0)}} & \frac{\beta}{q_{(N-1,0)}} \\
                               &                        &                              &                          & \frac{m\mu}{q_{(N,0)}}
    \end{array}\right),
    \\
    {\bf C}_{0}(0) &=& \left(\begin{array}{cccccc}
        c'_{(1,0)}   & c''_{(1,0)}  &      &      & & \\
        c_{(2,0)}    & c'_{(2,0)}   & c''_{(2,0)} &      & & \\
                     & c_{(3,0)}    & c'_{(3,0)}  & c''_{(3,0)} & & \\
                     &              & \ddots      & \ddots & \ddots & \\
                     &              &             & c_{(N-1,0)} & c'_{(N-1,0)} & c''_{(N-1,0)}
    \end{array}\right),
\end{eqnarray*}
where $c_{(i,0)}=q^{-1}_{(i,0)}(i-1)(\gamma+\tau_1+\tau_2+(1-m)\mu)$, for $i\in\{2,...,N\}$, $c'_{(i,0)}=q^{-1}_{(i,0)}((N-i)(1-m)\mu+(i-1)m\mu)$, for $i\in\{1,...,N\}$, and $c''_{(i,0)}=q^{-1}_{(i,0)}(N-i)((i-1)\beta+m\mu)$, for $i\in\{1,...,N-1\}$.
\section{Expressions for matrices ${\bf B}_{j}$, ${\bf C}_{j}(0)$, ${\bf C}_{j-1}(1)$ and ${\bf C}_{j+1}(2)$ in Eq. (6)}
\label{app:B}
In Eq. (\ref{eq:4}), ${\bf B}_{j}$ and ${\bf C}_{j}(0)$ are square matrices of order $N-j$, and matrices ${\bf C}_{j-1}(1)$ and ${\bf C}_{j+1}(2)$ are of dimension $(N-j)\times (N-j+1)$ and $(N-j)\times (N-j-1)$, respectively. They are given by
\begin{eqnarray*}
    {\bf B}_{j} &=& \left(\begin{array}{ccccc}
        \frac{m\mu}{q_{(j+1,j)}} & \frac{(N-j-1)\beta}{q_{(j+1,j)}} &                        &        & \\
                               & \frac{m\mu}{q_{(j+2,j)}} & \frac{(N-j-2)\beta}{q_{(j+2,j)}} &        & \\
                               &                        & \ddots                       & \ddots & \\
                               &                        &                              & \frac{m\mu}{q_{(N-1,j)}} & \frac{\beta}{q_{(N-1,j)}} \\
                               &                        &                              &                          & \frac{m\mu}{q_{(N,j)}}
    \end{array}\right),
    \\
    {\bf C}_{j}(0) &=& \left(\begin{array}{cccccc}
        c'_{(j+1,j)}   & c''_{(j+1,j)}  &      &      & & \\
        c_{(j+2,j)}    & c'_{(j+2,j)}   & c''_{(j+2,j)} &      & & \\
                       & c_{(j+3,j)}    & c'_{(j+3,j)}  & c''_{(j+3,j)} & & \\
                       &                & \ddots        & \ddots & \ddots & \\
                       &                &               & c_{(N-1,j)} & c'_{(N-1,j)} & c''_{(N-1,j)}
    \end{array}\right),
    \\
    {\bf C}_{j-1}(1) &=& \left(\begin{array}{ccccc}
        \frac{j(\gamma+\tau_2+(1-m)\mu)}{q_{(j+1,j)}} & \frac{jm\mu}{q_{(j+1,j)}} & & & \\
        & \frac{j(\gamma+\tau_2+(1-m)\mu)}{q_{(j+2,j)}} & \frac{jm\mu}{q_{(j+2,j)}} & & \\
        & & \ddots & \ddots & \\
        & & & \frac{j(\gamma+\tau_2+(1-m)\mu)}{q_{(N-1,j)}} & \frac{jm\mu}{q_{(N-1,j)}}
    \end{array}\right),
    \\
    {\bf C}_{j+1}(2) &=& \left(\begin{array}{cccc}
        \frac{(N-j-1)j(1-c)\beta}{q_{(j+1,j)}} & & &  \\
        & \frac{(N-j-2)j(1-c)\beta}{q_{(j+2,j)}} & &  \\
        & & \ddots &  \\
        & & & \frac{j(1-c)\beta}{q_{(N-1,j)}} \\
        & & & 0
    \end{array}\right),
\end{eqnarray*}
where $c_{(j+i,j)}=q^{-1}_{(j+i,j)}(i-1)(\gamma+\tau_1+\tau_2+(1-m)\mu)$, for $i\in\{2,...,N-j\}$, $c'_{(j+i,j)}=q^{-1}_{(j+i,j)}((N-j-i)(1-m)\mu+(i-1)m\mu)$, for $i\in\{1,...,N-j\}$, and $c''_{(j+i,j)}=q^{-1}_{(j+i,j)}(N-j-i)((i-1)\beta+m\mu)$, for $i\in\{1,...,N-j-1\}$.
\section{Expressions for matrices ${\bf E}_{j}$, ${\bf F}_{j}$, ${\bf D}_{j}(0)$, ${\bf D}_{j-1}(1)$ and ${\bf D}_{j+1}(2)$ in Theorems 2 and 3}
\label{app:C}
\vspace*{12pt}
Matrices ${\bf E}_{j}$, for $j\in\{1,...,N\}$, and ${\bf F}_{j}$, for $j\in\{1,...,N-1\}$, in Theorems 2 and 3 are of dimension $(N-j+1)\times (N-j+1)$ and $(N-j+1)\times (N-j)$, respectively, and are given by
\begin{eqnarray*}
    {\bf E}_{j} &=& \left( \begin{array}{cccc}
        \frac{m\mu}{q_{(j,j)}} & & & \\
                               & \frac{m\mu}{q_{(j+1,j)}} & & \\
                               &                          & \ddots & \\
                               &                          &        & \frac{m\mu}{q_{(N,j)}}
    \end{array}\right),
    \\
    {\bf F}_{j} &=& \left( \begin{array}{cccc}
        \frac{(N-j)(1-c)\beta}{q_{(j,j)}} & & & \\
                                          & \frac{(N-j-1)(1-c)\beta}{q_{(j+1,j)}} & & \\
                                          &                                       & \ddots & \\
                                          &                                       &        & \frac{(1-c)\beta}{q_{(N-1,j)}} \\
                                          &                                       &        & 0
    \end{array}\right).
\end{eqnarray*}
\par Matrices ${\bf D}_{j}(0)$, ${\bf D}_{j-1}(1)$ and ${\bf D}_{j+1}(2)$ are of dimension $(N-j+1)\times (N-j+1)$, $(N-j+1)\times (N-j+2)$ and $(N-j+1)\times (N-j)$, respectively, and have the form
\begin{eqnarray*}
{\bf D}_{j}(0) &=& \left(\begin{array}{cccccc}
        d'_{(j,j)}   & d''_{(j,j)}  &      &      & & \\
        d_{(j+1,j)}  & d'_{(j+1,j)} & d''_{(j+1,j)} &      & & \\
                     & d_{(j+2,j)}  & d'_{(j+2,j)}  & d''_{(j+2,j)} & & \\
                     &              & \ddots        & \ddots    & \ddots & \\
                     &              &               & d_{(N-1,j)}   & d'_{(N-1,j)} & d''_{(N-1,j)} \\
                     &              &               &               & d_{(N,j)}    & d'_{(N,j)}
    \end{array}\right),
\\
{\bf D}_{j-1}(1) &=& \left(\begin{array}{ccccc}
        \frac{(j-1)(\gamma+\tau_2+(1-m)\mu)}{q_{(j,j)}} & \frac{(j-1)m\mu}{q_{(j,j)}} & & & \\
                                                        & \frac{(j-1)(\gamma+\tau_2+(1-m)\mu)}{q_{(j+1,j)}} & \frac{(j-1)m\mu}{q_{(j+1,j)}} & & \\
                                                        &                                                   & \ddots                        & \ddots & \\
                                                        &                                                   &                               & \frac{(j-1)(\gamma+\tau_2+(1-m)\mu)}{q_{(N,j)}} & \frac{(j-1)m\mu}{q_{(N,j)}}
    \end{array}\right),
\\
{\bf D}_{j+1}(2) &=& \left(\begin{array}{cccc}
        \frac{(N-j)(j-1)(1-c)\beta}{q_{(j,j)}} & & & \\
                                               & \frac{(N-j-1)(j-1)(1-c)\beta}{q_{(j+1,j)}} & & \\
                                               &                                            & \ddots & \\
                                               &                                            &        & \frac{(j-1)(1-c)\beta}{q_{(N-1,j)}} \\
                                               &                                            &        & 0
    \end{array}\right),
\end{eqnarray*}
where $d_{(j+i,j)}=q_{(j+i,j)}^{-1}i(\gamma+\tau_1+\tau_2+(1-m)\mu)$, for $i\in\{1,...,N-j\}$, $d'_{(j+i,j)}=q_{(j+i,j)}^{-1}((N-j-i)(1-m)\mu+im\mu)$, for $i\in\{0,...,N-j\}$, and $d''_{(j+i,j)}=q_{(j+i,j)}^{-1}(N-j-i)(i\beta+m\mu)$, for $i\in\{0,...,N-j-1\}$.

\nocite{*}



\end{document}